\documentclass[traditabstract]{aa}

\usepackage{graphicx}
\usepackage{txfonts}
\usepackage{natbib}
\usepackage{xspace}
\usepackage{url}
\bibpunct{(}{)}{;}{a}{}{,}

\newcommand{\ho}{\ensuremath{H_{\rm{0}}}\xspace} 

\begin{document}

\title{COSMOGRAIL: the COSmological MOnitoring of \\GRAvItational Lenses}
\subtitle{XI. Techniques for time delay measurement in presence of microlensing}
\titlerunning{COSMOGRAIL XI -- Techniques for time delay measurement in presence of microlensing}

\author{M. Tewes \and F. Courbin \and G. Meylan}
\authorrunning{Tewes et al.}

\institute{Laboratoire d'astrophysique, \'Ecole Polytechnique F\'ed\'erale de Lausanne (EPFL), Observatoire de Sauverny, 1290 Versoix, Switzerland, \email{malte.tewes@epfl.ch}}

\date{Recieved 30 July 2012 / Accepted 21 February 2013}

\abstract{Measuring time delays between the multiple images of gravitationally lensed quasars is now recognized as a competitive way to constrain the cosmological parameters, and it is complementary with other cosmological probes. 
This requires long and well sampled optical light curves of numerous lensed quasars, such as those obtained by the COSMOGRAIL collaboration. High-quality data from our monitoring campaign call for novel numerical techniques to robustly measure the delays, as well as the associated random and systematic uncertainties, even in the presence of microlensing variations.
We propose three different point estimators to measure time delays, which are explicitly designed to handle light curves with extrinsic variability. These methods share a common formalism, which enables them to process data from $n$-image lenses. Since the estimators rely on significantly contrasting ideas, we expect them to be sensitive to different bias sources.
For each method and data set, we empirically estimate both the precision and accuracy (bias) of the time delay measurement using simulated light curves with known time delays that closely mimic the observations.
Finally, we test the self-consistency of our approach, and we demonstrate that our bias estimation is serviceable. These new methods, including the empirical uncertainty estimator, will represent the standard benchmark for analyzing the COSMOGRAIL light curves.}


\maketitle

\section{Introduction}

In the era of precision cosmology, in which a concordance model seems to fit independent observations, it is of utmost importance to both compare and combine all possible methods that constrain cosmological parameters. Comparing them yields an invaluable cross-check of the methods and the model. Combining them allows breaking the degeneracies inherent in single techniques. Probes including baryonic acoustic oscillations, weak lensing, supernovae, and cosmic microwave background measurements fit in this context exactly.

Also among these probes is the so-called ``time-delay method'', first proposed by \citet{Refsdal:1964vh} to measure cosmological distances independently of any standard candle. In practice, the method uses strongly lensed quasars with significant photometric variability. Photons emitted by the source quasar propagate towards us along different optical paths, resulting in multiple images. The light travel times associated to these images differ due to (i) the different path lengths, and (ii) the different Shapiro delays induced by the gravitational field of the lensing galaxy. As a consequence, the same quasar variability is seen with distinct time shifts in the light curves of the multiple quasar images. This paper presents methods of inferring the relative time delays between the quasar images, from such resolved, i.e. unblended, light curves.

Measured time delays, in combination with deep HST imaging and dynamical information on the lensing galaxy lead to competitive measurement of the Hubble constant \ho \citep[e.g.,][]{Suyu:2009ig, Suyu:2010fq}. The complementarity between quasar time delays and several other cosmological probes has been illustrated recently by \citet{Linder:2011cs}, who points out that the dark-energy figure of merit of a combination of Stage III experiments is improved by a factor of 5 if 150 quasar time delays are added. This also holds if the Universe is not assumed to be flat. It is noteworthy that adding this time delay information is very cheap compared to other Stage III or IV projects.

The COSMOGRAIL collaboration has now gathered almost a decade of photometric points for about 30 lensed quasars.
With such data, the time delays can in most cases be seen clearly ``by eye''. The data analysis is no longer about sorting out which time delay is the best among several plausible yet incompatible possibilities, but rather about performing an accurate measurement of the delay that can be reliably used for cosmology. New \emph{curve-shifting techniques} must be devised to extract the delays from such curves, which sometimes include a thousand points and typically display substantial microlensing variability due to stars in the lensing galaxy.

In this paper we present three independent curve-shifting algorithms that can deal with extrinsic variability. Our motivation behind the development of \emph{several} techniques is to provide a range of methods that rely on different principles. While the methods might not be free of systematics, we expect them to be biased in different ways, and we devote a large part of this work to estimating comprehensive error bars. 
Comparing the results from different curve-shifting techniques will allow us, in particular, to systematically cross-check our quantification of the biases.

Our paper is structured as follows: Section \ref{curves} gives an overview of features to be expected in light curves, most of them complicating the time delay extraction problem. We then present the point estimation formalism that is common to our curve-shifting techniques in Section \ref{techniques}. Sections \ref{splines} to \ref{disp} describe the three techniques, and we explain how we consistently compute error bars for each time delay and technique in Section \ref{errorbars}. 
We compare our techniques and the associated uncertainty estimates in Section \ref{test}, using a set of simulated light curves with known time delays. Finally, we present a summary and our conclusions in Section \ref{conclusions}.

\section{Light curves of lensed quasars}
\label{curves}

The COSMOGRAIL monitoring program obtains decade-long light curves from 1-2 m class telescopes \citep[e.g.,][]{Courbin:2011bl}. This data is reduced in a homogeneous way using deconvolution photometry. In the following we enumerate the properties and effects that will or might be observed in light curves from such ground-based optical observations.

\begin{enumerate}

\item {\bf Sampling and season gaps:} the data have irregular sampling, spaced on average by three to four days for typical COSMOGRAIL curves. By construction of the light curves, all quasar images of a lensing system are observed at the same epochs. The sampling can show some amount of periodicity on the scale of one day, since the targets tend to always be observed at optimal airmass. Almost all light curves are also affected by gaps of two to five months, corresponding to a period of the year where the lens is not observable.

\item {\bf Time delays:} by definition, the time delays produce a time-shifted version of the original variability curve of the source quasar. These delays range from hours to years. Depending on the length of the delays and the nonvisibility gaps, the intrinsic light curve of the quasar source may be fully or only partially sampled by the observations. The size of the ``overlap'' periods, in which several quasar images follow the same intrinsic source variability, strongly varies from one lens to another. In the worst case, for delays of roughly half a year (modulo one year), those variability features that are well observed in the light curve of one quasar image will be missed in the light curve of the other image, due to the nonvisibility gaps. This certainly exacerbates, and sometimes prohibits, measuring the delay.

\item {\bf Macrolensing image flux ratios:} the different strong lensing magnifications, as well as possible absorption by the lens galaxy or lensing perturbations by its satellites, yield stationary flux ratios between the lensed quasar images. Since the light curves are usually manipulated in magnitude scales, this translates into magnitude shifts between the curves.

\item {\bf Variable microlensing:} stars moving in the lensing galaxy act as secondary lenses that induce independent flickering of the light curve of each image. While this effect is interesting in itself as a tool to zoom on the lensed quasar and/or to probe the mass of these microlenses (see, e.g., \citealt{Refsdal:2000tm}, \citealt{Wambsganss:2000uc}, \citealt{Eigenbrod:2008dw}, \citealt{Eigenbrod:2008iq} and, for a short review, \citealt{Kochanek:2007tz}), it is a large complication in time delay determinations. Microlensing variations can occur on a broad range of time scales. We refer to \emph{slow} microlensing when speaking about any extrinsic variability that happens on time scales that are significantly larger than the intrinsic variability of the quasar. In extreme cases, microlensing can dominate the intrinsic variability of the quasar on all observable scales, preventing us from measuring time delays \citep[e.g.,][]{Morgan:2012cl}.

\item {\bf Variable source structure:} the light magnification by microlensing depends on source structure and size, i.e., micro-caustics due to stars may occur on spatial scales comparable in size to the quasar. In other words, in each source image, microlensing predominantly magnifies different parts of the source. As the total intrinsic luminosity of the quasar fluctuates, the light-emitting region might physically change in shape and size on the same time scales. This can introduce a mismatch between the light curves, which correlates with the intrinsic variability of the quasar or motion of its components \citep{Schechter:2003ep}. In particular, intrinsic variability patterns might be seen with different amplitudes in the light curves \citep[see][also for a curve-shifting method that tackles this issue]{Barkana:1997cv}.

\item {\bf Spurious additive flux:} the photometry of the quasar images might suffer from light contamination by the lensing galaxy or by the lensed images of the quasar host galaxy, resulting in constant additive shifts in \emph{flux} (not in magnitude) in the light curves. If in addition the photometric points are obtained from different telescopes or instruments, i.e., different resolutions, filters, and CCDs, these flux shifts might well be different for each setup. 

\item {\bf Flux sharing:}  this occurs in narrow blends of quasar images. The effect is due to the limited ability of photometric methods to separate the flux of individual images in such blends: while the total flux of the blend is measured very well, one observes random transfer of flux between the components, leading to negatively correlated scatter in the light curves. This problem is accentuated by bad seeing conditions.

\item {\bf Photometric calibration errors:} positively correlated scatter between the light curves, owing to noise/inaccuracies in the photometric normalization, i.e. magnitude zero point, of each observing epoch. This normalization is carried out using stars in the field of view. Small variability of some of the considered stars, as well as color terms that are unaccounted for, can contribute to errors in the relative flux calibration of the CCD frames. This correlated noise, and also the negatively correlated noise described under point 7, are particularly problematic when attempting to measure time delays that are shorter than the typical light curve sampling intervals.

\end{enumerate}
None of these effects is anything new. They affect all past and present optical monitoring programs. However, their significance increases with the quality of the data.

When considering only points 1 to 3 above, the problem of extracting time delays from noisy light curves is easy to formulate, as it literally corresponds to ``curve shifting'' along the time and magnitude axes. A wide variety of methods have been proposed to tackle the problem, from cross-corelations to simultaneous model fits. \citet{Hirv:2011wt} provide an overview of the different existing approaches, and present an algorithm based on the optimal prediction technique by \citet{Press:1992jj}. Recent works focusing on the statistical tools include a Bayesian estimation scheme \citep{Harva:2008wy} and a kernel-based approach combined with an evolutionary algorithm \citep{CuevasTello:2010td}.

Only a few of the existing techniques address the problem of extrinsic variability due to microlensing, or at least acknowledge this variability in their time delay uncertainty estimation. This can be attributed to the lack of long light curves of high enough quality to clearly exhibit extrinsic variability. If incorporated, models for microlensing variability were kept very simple (e.g., linear trends). A notable exception is the method of \citet{Morgan:2008fc}, which uses a \emph{physical} microlensing model in a Bayesian formalism \citep{Kochanek:2004ir}. However, the latter has a high computational cost, prohibitive in the case of decade-long curves of quadruply lensed quasars.

The methods presented in the following sections share a pragmatic approach to the mathematical representation of extrinsic variability. In this paper we are not interested in any modeling of microlensing but instead want to minimize and estimate its effect on the time delay measurement. Our methods should not be used to evaluate the odds of mutually exclusive time delay measurements. They are designed to accurately measure delays within narrow ranges around uncontroversial approximative solutions.

Inconsistent delay measurements often result from insufficient data. A noticeable example of such a situation is the decade-long controversy about two competing values of the time delay of Q~0957+561 \citep{Walsh:1979cz}, measured in the optical ($\Delta t = 415 \pm 20$ days) by \citet{Vanderriest:1989uj} and in the radio ($\Delta t = 513 \pm 40$ days) by \citet{Lehar:1992bg}. The debate was closed by \citet{Kundic:1997br} using additional photometric measurements, as opposed to refined methods. A more recent example is the delay measurement of HE~1104-1805 \citep[see, e.g.,][]{GilMerino:2002fi, Pelt:2002ke}. In our opinion the reliability of time delay measurement techniques has often been overestimated, especially when insufficient data could not hint at any potential extrinsic variability.


\section{The time-shift formalism}
\label{techniques}

All three curve-shifting techniques presented in this paper are \emph{point estimators}; they can be seen as functions that take $n$ light curves as input ($n \ge 2$, usually $2$ or $4$) and return $n$ corresponding \emph{time shifts} $\tau$, one for each curve. These optimal time shifts minimize the mismatch between the common intrinsic quasar variability features of all the light curves. \citet{Pelt:1998vc} follow a similar approach, but attribute a shift of zero to an arbitrarily chosen curve. Provided that the optimization algorithm is very robust, such as a brute force exploration, this is equivalent to our choice of adjusting all $n$ time shifts. The individual optimal time \emph{shifts} are not informative, but they directly translate into unambiguous time delay estimations between each pair of curves:
\begin{equation}
\label{defdelay}
\Delta t_{\mathrm{XY}} = \tau_{\mathrm{Y}} - \tau_{\mathrm{X}}.
\end{equation}
As a result, $n$ time-shift estimates simultaneously obtained from $n$ light curves of a lens system yield $n(n-1)/2$ dependent -- and consistent -- time delay estimations. Note that this would naturally generalize to probability distributions instead of point estimates.

By construction, this trivial time-shift formalism avoids selecting any reference curve with respect to which $n$ independent delays would be expressed. This is crucial since it can well be that, for example, strong extrinsic variability in quasar image $\mathrm{A}$ prevents us from measuring the delays $\Delta t_{\mathrm{AB}}$ and $\Delta t_{\mathrm{AC}}$, while the delay $\Delta t_{\mathrm{BC}}$ can be well determined, as observed in the case of HE~0435-1223 \citep{Courbin:2011bl}.
Furthermore, methods complying with this formalism shift all four light curves of quad lenses simultaneously. This is a strong advantage especially in the presence of extrinsic variability, because using all curves constrains the intrinsic variability much better than a pairwise processing.

Important is that we also exploit the common formalism of our point estimators by estimating their variance and bias in exactly the same way (Section \ref{errorbars}). Before describing the three methods, we underline that our point estimators all rely on iterative nonlinear optimization algorithms. As a consequence, they all show a certain amount of dependence upon the choice of initial guesses for the time shifts. For each lens system to be analyzed, we systematically evaluate this dependence by running our methods a few hundred times on the exact same observed light curves, starting from initial shifts randomly selected in a range of generally $\pm 10 $ days around a plausible solution. We call the variance of the resulting monomodal distributions of time delays the \emph{intrinsic variance} of a delay estimator applied to the particular set of curves. We illustrate this in Section~\ref{splines}. In principle, this intrinsic variance can be made arbitrarily small, by increasing the robustness and precision of the optimizations. In practice, a compromise with CPU cost has to be found. We have implemented our methods so that their intrinsic variance is significantly smaller than the other sources of error. In any case, the intrinsic variance will be part of the total uncertainty evaluation. Furthermore, we always use the mean of these distributions as our best time delay estimations between observed light curves.


\section{Method 1: simultaneous spline fit}
\label{splines}

\begin{figure*}
\resizebox{\hsize}{!}{\includegraphics{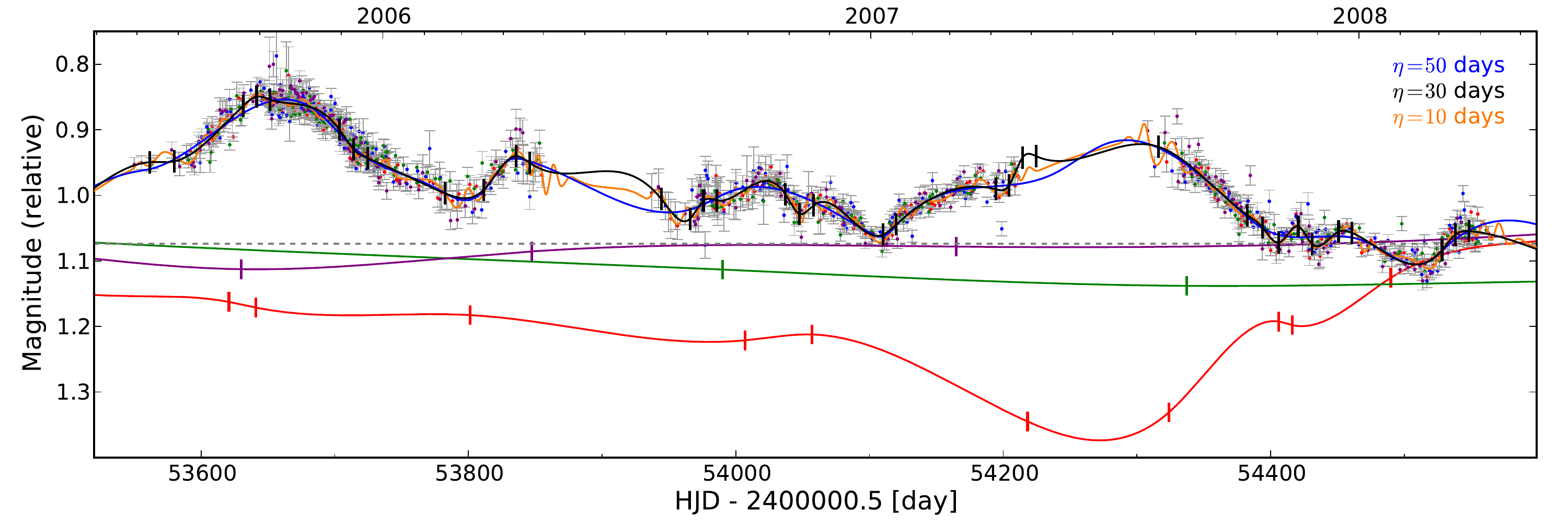}}
\caption{Illustration of the free-knot spline technique. The data points are shifted COSMOGRAIL light curves of the quadruply lensed quasar HE~0435-1223, published in \citet{Courbin:2011bl}. A reasonable spline representing the intrinsic variability is shown in black (initial knot step $\eta = 30$); knots are shown as vertical ticks. The red, green, and violet splines represent the relative extrinsic variability corrections, applied to the observed light curves A, C, and D (respectively), so that they match the common intrinsic spline. These extrinsic splines are plotted relative to the dashed gray line. The optimization starts with uniformly distributed knots, and the knots tend to migrate to areas in which they are required by well-constrained patterns of the data. The blue and orange curves are alternative intrinsic splines (shown without knots to avoid clutter) with inadequate knot densities of $\eta = 50$ and $10$ (see text).}
\label{fig_spline1}
\end{figure*}

Our first method fits a single continuous model to all data points of the light curves, simultaneously adjusting time and magnitude shifts between these curves so as to minimize a $\chi^2$ fitting statistic between the data points and the model. We designate this common model as \emph{intrinsic}, even when microlensing might prevent us in practice from getting access to the pure intrinsic variability of the source quasar. The idea of fitting such a single model to shifted light curves defines a whole family of existing time delay measurement methods. These techniques differ by the mathematical representation of the intrinsic curve; for instance, \citet{Press:1992jj} use a Gaussian process, \citet{Lehar:1992bg} use Legendre polynomials, \citet{Barkana:1997cv} uses cubic splines with equidistant knots, \citet{Burud:2001ka} use a regularized numerical model, \citet{CuevasTello:2006fn} use a linear combination of Gaussian kernels, and \citet{Vakulik:2009br} use a linear combination of sinc functions.

In the presence of independent ``extrinsic'' variability such as slow quasar microlensing, light curves will not adequately overlap for any shifts in time and magnitude. It is easy to conceive that a mismatch of this type can lead to strongly biased time delay estimations. It is therefore mandatory to explicitly model the extrinsic variability, at the price of an increased number of free parameters. For example, to represent the microlensing in their high-quality light curves of the lensed quasar HE~0435-1223, \citet{Kochanek:2006fp} use independent quadratic polynomials for each season.

Models representing the relative extrinsic variability act similarly to high-pass filters on the data. If these models are as flexible as the intrinsic curve, they can compensate for incorrect time shifts, and the information of the time delay is lost. This is our main motivation in proposing a new method that tries to \emph{locally} adapt the flexibility of the models to the peculiarities of the curves.

\subsection{Free-knot splines: the principle}
\label{bok}

We use so-called \emph{free-knot} B-splines to represent both the intrinsic and the extrinsic variability of the light curves. A spline is a piecewise polynomial function, and its knots are the locations where the polynomial pieces connect. We only consider splines of degree 3, i.e., those with continuous second derivatives all across the curves. For these free-knot regression splines, not only the polynomial coefficients but also the knot positions are seen as free variables to be optimized. These splines yield significantly better fits than uniform splines, for a given number of knots. Furthermore, they do not introduce any arbitrary discrete grid in the model, which is of high importance for our application. Ideally we want our models to be shift-invariant in terms of their ability to fit any given pattern.

In the case of fixed knots, finding the unique least-squares spline approximation to some data points is a linear problem. This property does not hold for free-knot splines: optimizing the knot positions to minimize the $\chi^2$ requires nonlinear parameter estimation. The nonlinear optimization is particularly difficult, since the motion of the knots leads to many local optima and stationary areas in the parameter space.

\citet{Molinari:2004tv} present an efficient algorithm named ``bounded optimal knots'' (BOK) to optimize the knot locations of least-squares spline approximations. The authors recall that fitting a free-knot spline is a problem that can be separated in a linear and a nonlinear part \citep{Golub:1973wb}. For any given knot configuration, the computation of the corresponding optimal spline \emph{coefficients} remains linear, hence fast. For this task, our implementation makes use of wrappers around FITPACK \citep{Dierckx:1995tp} provided by \verb+scipy+\footnote{Scientific Tools for Python \citep{numpyscipy},\\ \url{http://www.scipy.org/}}. The main idea of BOK is to wrap this linear coefficient computation inside an iterative \emph{bounded} optimization of the knots.
The bounded optimization guarantees a well-defined minimal distance between the knots, by keeping them confined to disjoint windows. This scheme avoids the ``coalescence'' (i.e., superposition) of knots, which would correspond to unwanted discontinuities in the derivatives of the spline. Following an idea of the \emph{Evolutionary BOK} algorithm \citep[also described by][]{Molinari:2004tv}, we update the bounds of the windows once the knot locations have been robustly optimized, and iteratively repeat the process. This mechanism effectively moves the windows to follow their knots, yet always ensuring a minimal knot distance.

Fitting a single spline to fixed data points is only a fragment of the curve-shifting problem: our model for the light curves not only consists of a common intrinsic spline, but also of several independent extrinsic splines. In addition, to make our curve-shifting technique efficient, we adjust the time shifts between the curves simultaneously with the splines, instead of performing independent fits for different trial delays. In mathematical terms, we aim at finding the global minimum of
\begin{equation}
\chi^2 = \sum_{i=1}^{n} \sum_{j=1}^{N_i} \frac{[ m_{ij} - s(t_{ij} + \tau_{i}) -  \mu_{i}(t_{ij})]^2}{\sigma_{ij}^2},
\end{equation}
in which $n$ is the number of light curves with $N_i$ photometric points $(t_{ij}, m_{ij} \pm \sigma_{ij})$, $\tau_i$ are the time shifts, $s$ is the intrinsic spline, and  $\mu_{i}$ are the extrinsic splines.

We minimize this $\chi^2$ in an iterative process in which the splines and the time shifts get optimized one after the other, using custom strategies for each parameter. Several formal difficulties arise as the ``footprint'' of the mixed data points evolves when the time shifts are modified; for instance, we have to stretch the knot locations so that they follow the extent of a spline's domain. For details of the admittedly intricate but modular optimization procedure, we refer the reader to the documentation accompanying the source code.

\subsection{Free-knot splines in practice}

\begin{figure*}
\resizebox{\hsize}{!}{\includegraphics{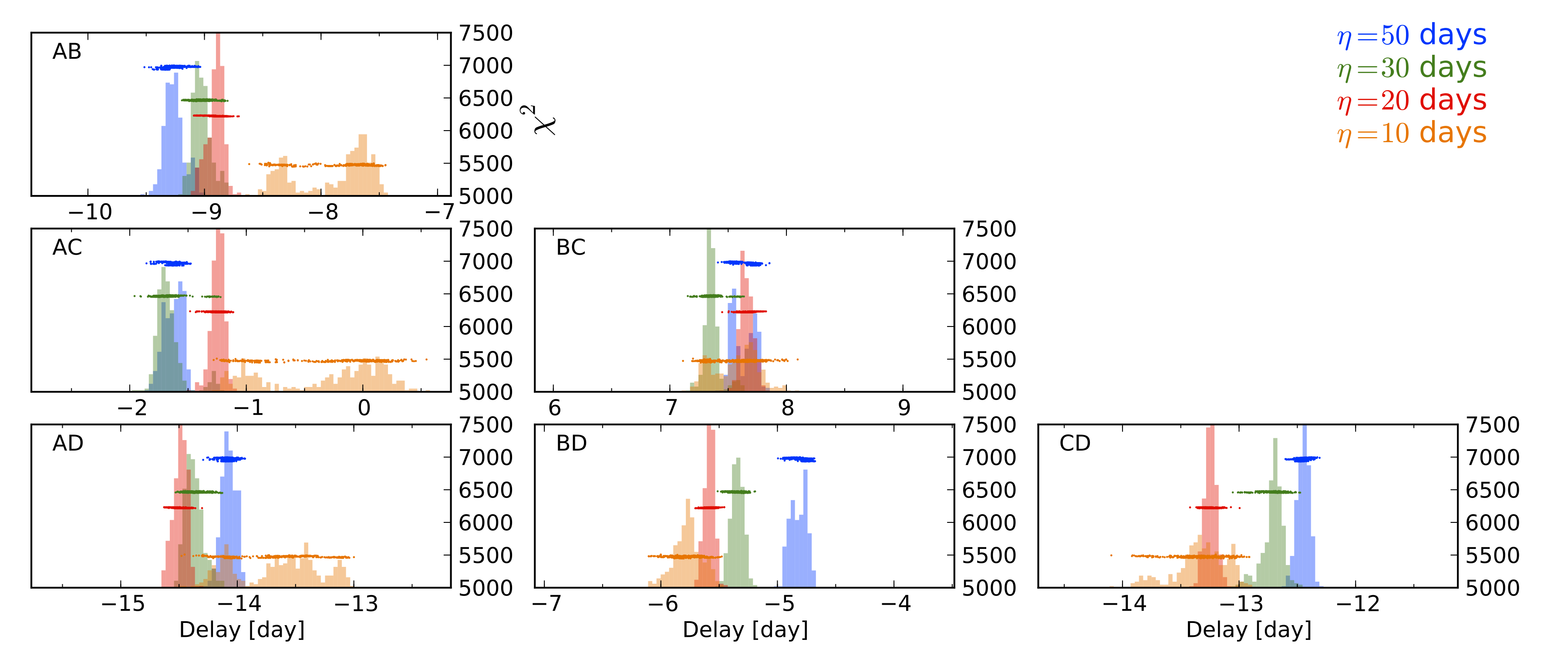}}
\caption{Distributions of time delays obtained by running the free-knot spline technique always on the \emph{same} data, starting from randomized initial time shifts. The light curves are from \citet{Courbin:2011bl}, an excerpt of which is shown in Fig. \ref{fig_spline1}. The colors encode the initial knot step $\eta$ of the intrinsic spline, spanning a wide range corresponding to a factor 5 in the number of knots. Superposed on the histograms are scatter plots of the minimal $\chi^2$ values obtained by the optimization. We stress that the histograms shown here are not to be mistaken for probability density functions of time delay measurements on HE~0435-1223. They only illustrate the intrinsic variance, i.e., the finite precision of the optimization algorithm, as applied to a high-quality data set.}
\label{fig_spline2}
\end{figure*}

In Fig. \ref{fig_spline1}, we show three seasons of a spline fit obtained for the light curves of the quadruply lensed quasar HE~0435-1223 \citep{Courbin:2011bl}. Variants of the intrinsic spline are drawn on top of the shifted data points, while the red, green, and purple splines represent the extrinsic variability for which the data points have been corrected (see caption).

Each spline is parametrized by the knot epochs, as well as by the associated coefficients. The curve shifting starts with equidistant knots, and flat splines. Before running the optimization, one has to choose a number of knots for the intrinsic spline (e.g., one knot every 30 days, $\eta = 30$), the number of knots for each extrinsic spline (e.g., one knot every 150 days, depending on the apparent microlensing variability in each individual curve), and the respective minimum knot distances, $\epsilon$ (e.g. 10 days). These numbers remain unchanged by the fitting. They control the global flexibility of each spline.

How sensitive is the technique to these choices ? In Fig. \ref{fig_spline2} we illustrate the effect of the initial uniform knot step $\eta$ of the intrinsic spline on the resulting time delay measurements; a smaller step corresponds to more knots. These time delay histograms are obtained by repeatedly running the optimization on the \emph{same} data, starting from random initial time shifts, uniformly distributed within a range of $\pm10$ days around the typical best-fitting solutions. We also randomly shuffle the order in which the optimizer processes the microlensing splines, to marginalize over any possible asymmetry introduced by our iterative spline fitting algorithm.
For each number of knots, the distributions of measured time delays display a finite scatter. This is the \emph{intrinsic variance} introduced in Section \ref{techniques}. It depends on the robustness of the $\chi^2$ minimization algorithm, and thus also on the complexity (degeneracies, local minima) of the $\chi^2$ hypersurface. As we observe, the result of our optimization is by no means global. A visual inspection of the splines reveals that the knots tend to settle in a few different repeating configurations. This intrinsic variance will naturally contribute to the error bar that we attribute to a time delay measurement.

The influence of the step $\eta$ on the centroids of the measured delays in Fig. \ref{fig_spline2} is surprisingly small, considering that the range of tested values covers a factor 5 in the number of knots. Visualizing the intrinsic spline fits (3 examples in Fig \ref{fig_spline1}), an initial step of $\eta = 50$ days is clearly too large for this particular data set, as the resulting spline is not able to follow obvious intrinsic trends. On the other hand, setting $\eta = 10$ days yields far too many knots and the spline tends to fit the noise of individual curves. As expected, we observe in Fig. \ref{fig_spline2} that deliberately selecting too large a number of knots leads to an increase in the intrinsic variance of the method. This behavior is easily reproduced when changing the knot density of the extrinsic splines. Too few knots bias the measured time delays, too many knots ``dilute'' them, due to the degeneracies with the intrinsic spline. It is very possible to attribute an extrinsic spline to \emph{each} light curve instead of leaving one curve as a reference; this leads to obvious degeneracies between the extrinsic splines, but does not systematically increase the intrinsic variance.

Finally, we note that the choice of the minimum knot distance $\epsilon$ of the intrinsic spline, in a range from 2 to 15 days, has practically no effect at all on the time delay measurements. Only a very low minimal distance ($ \epsilon < 5$), combined with a high number of knots ($\eta < 15$) significantly increases the intrinsic variance of the delay measurements, without introducing systematic shifts. In all the following, we use a minimal knot distance of ten days. 

To close this section, we compute the number of free parameters involved in this technique. Every spline has a number $n_k$ of so-called \emph{internal} knots, i.e., knots located strictly within the temporal range of the data points. These are the knots that are free in case of a free-knot spline. For given internal knot epochs, a cubic spline is defined by $n_k + 4$ independent coefficients \citep[see, e.g.,][]{Molinari:2004tv, deBoor:1978wq}. These can be seen as one coefficient per internal knot, plus two coefficients for the knots located at each extremity. The only other free parameters are the time shifts $\tau$ of the curves. Let us consider the case of a quad lens with 500 monitoring epochs spread over 6 seasons, $n_k=75$ knots for the intrinsic spline, and $n_k=15$ knots for each of the 4 extrinsic splines. This corresponds to 2000 data points, and $(2 \times 75 + 4) + 4\,(2 \times 15 + 4) + 4 = 294$ free parameters. Our implementation of this simultaneous fit converges in less than a minute on a single ordinary CPU.

We postpone the assessment of the total uncertainties of delay measurements to Section \ref{errorbars}.

\section{Method 2: variability of regression differences}
\label{regdiff}

\begin{figure*}
\resizebox{\hsize}{!}{\includegraphics{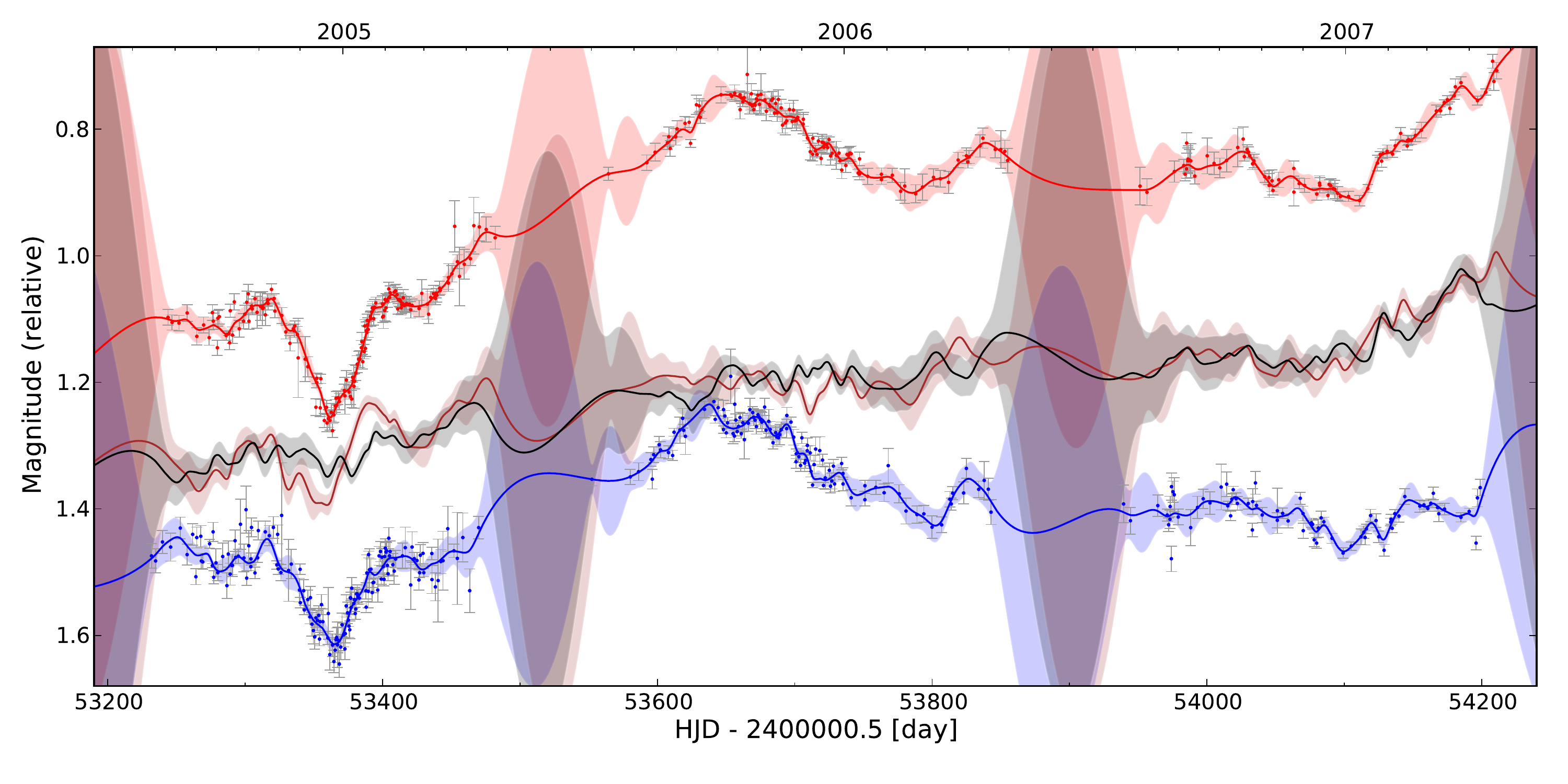}}
\caption{Illustration of the regression difference technique on light curves of HE~0435-1223 \citep{Courbin:2011bl}. For clarity, only the light curves of images $A$ (red) and $B$ (blue) are shown. The Gaussian process regressions are shown as red and blue continuous mean functions and $1\sigma$ envelopes. Curve B has been shifted in time with respect to A to minimize the weighted total variation of the $A - B$ difference curve, shown in black. The brown difference curve was obtained by deliberately shifting $B$ by 15 days with respect to its optimal position. The curves have been arbitrarily offset in magnitudes for display purposes.}
\label{fig_regdiff1}
\end{figure*}

Our second curve-shifting technique consists of a much less parametric approach, and can be summarized as follows.
\begin{enumerate}
\item Instead of simultaneously fitting one intrinsic model to all light curves, we start by performing an \emph{independent} regression on each individual curve. 
We can then easily express numerical \emph{difference curves} between time-shifted pairs of these finely sampled regressions. Since we work with light curves in magnitudes, these difference curves correspond to flux ratios.

\item The regression curves are shifted along the time axis to minimize the \emph{variability}, i.e., structures, of the difference curves. In mathematical terms, we propose to measure this variability through the ``weighted average variation'', a concept inspired by the \emph{total variation}. This approach minimizes the derivative of the difference curves, as opposed to the difference curves themselves.

\end{enumerate}

This method is illustrated in Fig. \ref{fig_regdiff1}. Regressions of the light curves of two quasar images are shown in red and blue. If the regressions are shifted in time to correctly compensate for the delays, any intrinsic variability pattern of the quasar cancels out in the difference light curves (black curve of Fig. \ref{fig_regdiff1}). In this particular situation, the difference curves contain only the relative extrinsic variability between the curves. If the absolute extrinsic variability is independent for each curve, any statistical property of the relative extrinsic variability between the curves will a priori be independent of the time shifts. In practice this is not entirely the case, owing, e.g., to the finite length of the light curves, which is much shorter than the largest time scales of microlensing variability.

In contrast, if the regressions are not shifted by the correct time delays, the intrinsic variability does not cancel out. As a consequence, the difference curves also contain a finite difference of the intrinsic quasar variability. If the latter is fast and strong enough, this corresponds to clear additional irregularities in the difference curves. These can be observed in the difference curve shown in brown on Fig. \ref{fig_regdiff1}, which was obtained by attributing incorrect time shifts to the regressions.

\subsection{Gaussian process regression difference curves}

The purpose of the regression is to allow the measurement of relative variability between time-shifted light curves, despite the highly irregular sampling and season gaps. To adequately weight localized variability according to its uncertainty, we need a regression that expresses not only a most likely value but also a confidence interval for this value at each epoch.

Gaussian process regression (GPR) is a powerful formalism whose predictor curve -- a Gaussian process -- does not follow a predetermined parametrized form. Instead, the regression curve is constrained in its freedom by a covariance function that describes how correlated two given points of the curve are, depending on their separation along the time axis. Given (i) such a prior covariance function, (ii) a prior for the regression function itself, and (iii) the observed data, the GPR yields a Gaussian distribution (i.e., a mean value and a variance) for the regression value at any interpolation epoch. For a pedagogical introduction to GPR, see e.g. \citet{Press:2007tn}.

To implement our curve-shifting technique, we make use of the GPR functionality provided by the \verb+pymc+\footnote{\url{http://pypi.python.org/pypi/pymc/}} python package \citep{Patil:2010tx}. Before computing a regression, we have to choose priors for both the covariance and a mean function. For the latter, we simply use an uninformative constant function, at the mean magnitude of the curve's data points. The choice of a prior covariance function is less trivial and certainly not unique. Several families of covariance functions are implemented in \verb+pymc+; in all the following we make use of the Mat\'ern family, with an amplitude parameter of 2.0 magnitudes, a scale of 200 days, and a smoothness degree $\nu = 1.5$. We observe that this choice yields good results for several different COSMOGRAIL data sets, in terms of the empirical time delay uncertainties determined in Section \ref{errorbars}. But a priori, these parameters can be fine tuned individually for each lens system to be analyzed, using this same criterion. When experimenting with covariance functions, it is of particular importance to ensure that the season gaps are interpolated with adequately large variances.

In practice, for each of the $n$ light curves, we evaluate the GPR every 0.2 days. Given some trial time shifts, we express the $n\,(n-1)/2$ difference curves by subtracting linearly interpolated magnitudes of the shifted regression curves. Indeed, each pair of curves has to be considered only once; for the variability analysis, the difference curve $A-B$ yields the same result as $B-A$. In a similar way, the uncertainties at each epoch of the difference curves are obtained by summing the linearly interpolated variances. We proceed by quantifying the variability of the difference curves.

\subsection{Minimizing the weighted average variation}
\label{TV}

To measure the variability of a difference curve, we define a simple scalar statistic, that we refer to as the \emph{weighted average variation} (WAV). It corresponds to an average absolute value of the discrete derivative along the difference curve. We weight the terms of this average by the local uncertainty of the difference curve, and normalize this weighting to cancel the influence of the varying size of the overlapping regions.
For a difference curve $f(t_j)$ with variance $\sigma^2(t_j)$, and N regularly sampled points ($j = 1, \ldots, N$), the WAV is given by
\begin{equation}
\mathrm{WAV}(f) \mathrel{\mathop:}= \frac{\sum_{j=1}^{N-1} \left| \hat{f'}(t_j) \right| \cdot w(j) }{\sum_{j=1}^{N-1} w(j) }
\end{equation}
where
\begin{equation}
\hat{f'}(t_j) =  \frac{f(t_{j+1}) - f(t_{j})}{t_{j+1} \, - \, t_{j}}
\end{equation}
and the weights are
\begin{equation}
w(j) =  \frac{2}{  \sigma(t_{j}) + \sigma(t_{j+1}) }.
\end{equation}
The time shifts of the regression curves are optimized using a nonlinear technique that minimizes a single scalar objective function obtained by summing the WAV of all pairwise difference curves. Due to the low number of parameters ($n$ time shifts) compared to the free-knot spline technique, this optimization can be made very robust; i.e., it leads to a negligible sensitivity to initial conditions. The total computing time for a quad lens with 500 epochs in each curve is on the order of one minute on a single CPU. The GPR takes more than half of this time.

\section{Method 3: dispersion minimization}
\label{disp}

Our third curve-shifting method is broadly inspired by the dispersion techniques from \citet{Pelt:1996vy}. In \cite{Courbin:2011bl}, we have already applied this particular time delay estimator to the light curves of HE 0435-1223. It has also been used in \citet{Eulaers:2011hz}. A notable difference from classical dispersion techniques is that we make use of a simple linear interpolation to form pairs of predicted and observed points, instead of considering only pairs of closeby observed points.

The method consists of a nonlinear optimization of the time shift of each light curve to minimize a single scalar objective function that quantifies the ``mismatch''. \emph{Dispersion functions} are such objective functions that do not involve any model for the intrinsic variability, but only use the relative dispersion between the \emph{time-shifted points} of the light curves. As for the free-knot spline method (Section \ref{splines}), it is necessary to explicitly compensate for any slow extrinsic variability before evaluating the dispersion for given trial time shifts. 
We represent this extrinsic variability by low-order polynomials added to either the full curves or to individual seasons \citep[for an illustration, see Fig. 5 of][]{Courbin:2011bl}. These polynomials are optimized simultaneously with the time shifts to minimize the same dispersion function.

\begin{figure}
\resizebox{\hsize}{!}{\includegraphics{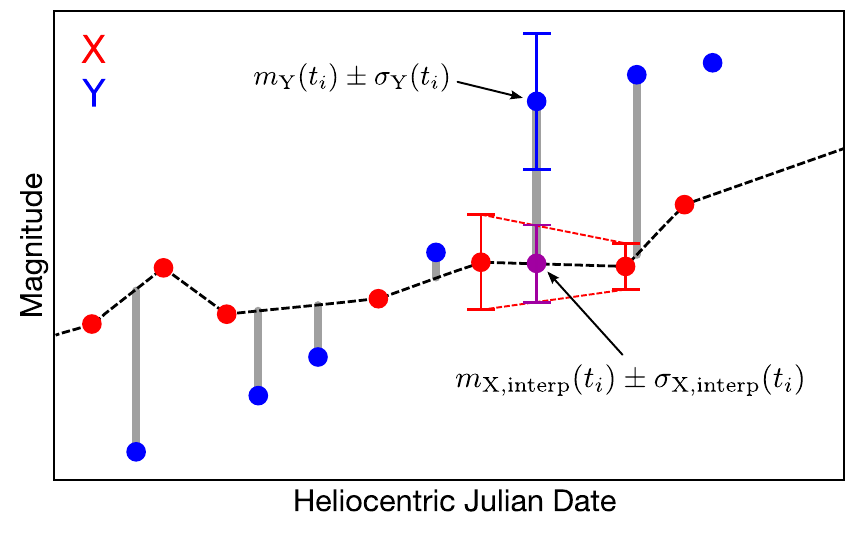}}
\caption{Illustration of the ``elementary'' dispersion function used in the curve-shifting technique described in Section \ref{disp}. The vertical gray bars represent the terms of the summation of equation \ref{equ_disp}. The last shown point of light curve Y would \emph{not} contribute to the dispersion, since it falls into a large gap of X. This elementary dispersion function is not invariant with respect to swapping the curves X and Y. However, our total dispersion estimate is symmetric, as we average these elementary dispersion across all permutations of 2 curves among $n$. Not shown in this sketch are the polynomial corrections for extrinsic variability. These corrections are optimized against the same total dispersion.}
\label{fig_disp}
\end{figure}

It remains to define the dispersion function. It should be able to evaluate the mismatch between $n \ge 2$ time-shifted light curves of a quasar lens, treating all those $n$ curves equally, i.e., without choosing an arbitrary reference curve. We achieve this by first defining an \emph{elementary} dispersion function that operates on pairs of curves, and then averaging its value over all $n(n-1)$ permutations of 2 among $n$ light curves. This elementary dispersion function between two curves X and Y is illustrated in Fig. \ref{fig_disp}. We linearly interpolate (and never extrapolate) the light curve X at each epoch of Y, provided that the time interval over which this interpolation is done is shorter than $\delta = 30 $ days. Given the typical sampling of light curves, this means that we do not interpolate across observing season gaps. We then compute a sum of square differences between the interpolated magnitudes $m_{\mathrm{X, interp}}(t_i)$ and the corresponding magnitude measurements $m_{\mathrm{Y}}(t_i)$ of light curve Y:
\begin{equation}
\label{equ_disp}
D^2(\mathrm{X}, \mathrm{Y}) = \frac{1}{N} \sum_{i=1}^{N} \frac{\left(m_{\mathrm{X, interp}}(t_i) - m_{\mathrm{Y}}(t_i)\right)^2}{\sigma_{\mathrm{X, interp}}^2(t_i) + \sigma_{\mathrm{Y}}^2(t_i)},
\end{equation}
where the summation goes over the $N$ epochs of Y for which the interpolation of X was performed, given the above criteria. Each term is weighted by a combination of the linearly interpolated uncertainty on $m_{\mathrm{X, interp}}(t_i)$ and the uncertainty of $m_{\mathrm{Y}}(t_i)$.

For a set $S$ of $n$ light curves, e.g. $S = \{\mathrm{A},  \mathrm{B}, \mathrm{C}, \mathrm{D} \}$, the average dispersion is computed by
\begin{equation}
\label{equ_totdisp}
D^2(\,S) = \frac{1}{n(n-1)} \sum_{\{\mathrm{X}, \mathrm{Y}\} \subset  \,S, \, X \neq Y} D^2(\mathrm{X}, \mathrm{Y}).
\end{equation}
This dispersion is a function of the time shifts and the extrinsic variability corrections of each light curve. In practice we minimize the average dispersion by using iteratively a simulated annealing optimizer for the time shifts and a Powell optimizer for the extrinsic variability, as implemented in \verb+scipy+ \citep[for a description of the algorithms, see e.g.][chap. 10.12 and 10.7]{Press:2007tn}.

Due to the sampling and scatter of the light curves, the dispersion function usually has a very rough structure in time shift space, resulting in a poorly defined minimum. It is tempting to ``smooth'' this dispersion function. This can be done, for instance, by using a regression instead of a simple linear interpolation, or by determining the dispersion minimum through a local fit \citep[see e.g. Fig 5. of][]{Vuissoz:2008bu}. Such measures clearly reduce the intrinsic variance (see Section \ref{techniques}) of the estimator by increasing its stability against random initial time shifts. But we also observe that they often introduce additional bias. We therefore prefer to simply use the dispersion function described above, tackling its ``noisy'' minimum with a robust optimization algorithm.

\section{Empirical estimation of time delay uncertainties}
\label{errorbars}

As described in Section \ref{techniques}, the three curve-shifting techniques presented in this paper can be seen as recipes that yield a single time \emph{shift} estimation $\tau$ for each light curve of a lensed quasar. In this section, we describe how we proceed to empirically obtain reliable error bars for time \emph{delays} between pairs of curves ($\Delta t_{\mathrm{XY}} = \tau_{\mathrm{Y}} - \tau_{\mathrm{X}}$), as measured through these point estimations $\tau$. This error analysis is done individually for each quasar lens, i.e., for each set of observed light curves. Indeed, our aim is to quantify how well the curve-shifting techniques perform on given particular data. We do not evaluate the overall capability of the techniques to measure time delays for various quasars or monitoring programs.

To compute the time delay uncertainties, we follow a Monte Carlo approach. We apply the curve-shifting techniques on a large number of synthetic light curves with \emph{known} time delays. This allows us to assess both the variance and the bias of our point estimators, as we relate in Section \ref{hists}. We draw these synthetic light curves from a comprehensive ``generative'' model, i.e., a mathematical model for randomly generating mock light curves from scratch, while controlling the hidden parameters such as the time delays. For a given lensed quasar, our generative model -- whose details are described in the following sections -- is composed of
\begin{enumerate}
\item A single intrinsic variability curve, common to all $n$ quasar images. We use the intrinsic free-knot spline obtained by applying our first curve-shifting method (Section \ref{splines}, black spline of Fig. \ref{fig_spline1}) to the observed light curves.
\item A smooth extrinsic variability curve (``slow'' microlensing) for each curve of the lens. Again, we directly use the curves obtained by the spline-fitting technique on the observed data (red, green, and purple splines of Fig. \ref{fig_spline1}).
\item An independent ``fast'' extrinsic variability, which can be seen as correlated noise, for each curve. This contribution is randomized, i.e., individually drawn for each realization of the synthetic curves.
\end{enumerate}

\begin{figure*}
\resizebox{\hsize}{!}{\includegraphics{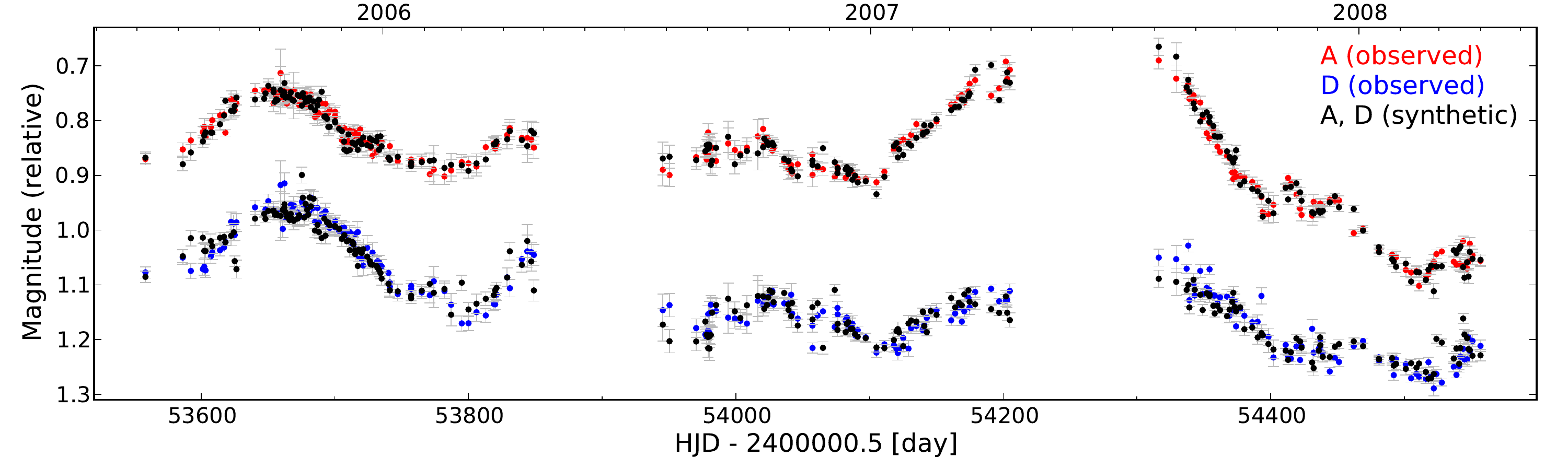}}
\caption{Synthetic curves (black) drawn from a well adjusted generative model (see text), and the corresponding observed data points for the lensed quasar HE~0435-1223. For illustration purposes only 3 seasons of 2 quasar images are shown, arbitrarily offset in magnitude. The purpose of the generative model is to simulate curves with \emph{known} time delays that nevertheless mimic the observations at best.}
\label{fig_comparawsims}
\end{figure*}

We always sample from this generative model at the actual observing epochs of the monitoring. As a result, we obtain synthetic curves that closely imitate the intrinsic and extrinsic variability, sampling, and scatter characteristics of the real data. If the properties of the randomized fast extrinsic variability are well adjusted, these synthetic curves are statistically undistinguishable from the observations. As illustrated in Fig. \ref{fig_comparawsims}, they could easily be mistaken for the real light curves. Nevertheless, they have known time delays.

\begin{figure*}
\resizebox{\hsize}{!}{\includegraphics{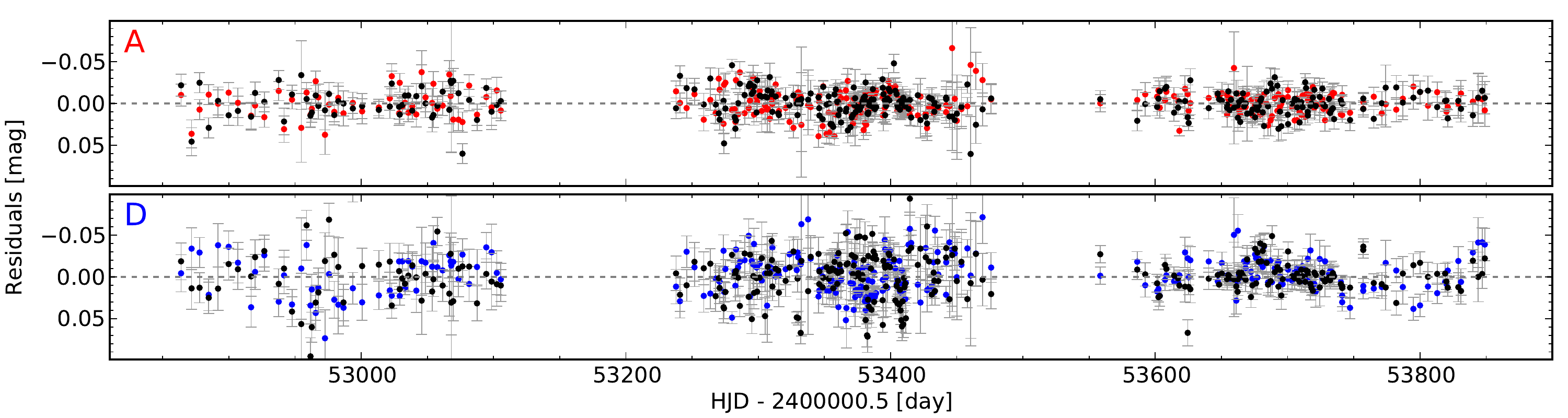}}
\caption{Residuals of the observed (red, blue) and the synthetic (black) light curve of images A and D of HE~0435-1223, as obtained by applying the free-knot spline method. The noise used for the synthetic curve is adjusted so that these black residuals show on average (i) the same standard deviation and (ii) the same number of ``runs'' as those of the observations. As described in the text, the power-law noise has been rescaled to locally match the scatter amplitude of the observed curves, as can be clearly seen in this figure.}
\label{fig_resi}
\end{figure*}

Our first objective of this methodology is to obtain synthetic curves that share a very similar ``time delay constraining power'' with the observations, so that we can assume our delay measurement methods to perform equally well or badly on both. Clearly, this constraining power of a set of light curves increases with the amount of intrinsic variability, and decreases as this variability gets diluted by extrinsic effects such as microlensing, sampling, and noise \citep{Eigenbrod:2005jx}. Furthermore, curve-shifting methods might not be able to optimally exploit the information content of the data points. In particular, they are likely to be \emph{biased} even by those peculiarities of the data that can be indubitably determined from the observations, such as the interplay between large-scale extrinsic and intrinsic variability, and season gaps. Our second objective in generating synthetic curves that mimic the real data as closely as possible is therefore to reveal these biases. Provided that a rough but unambiguous estimation of time delays can be made, the observed light curves do yield some nearly unmistakable information about the intrinsic and extrinsic variability. We use this information as a deterministic, smooth part of our generative model, and marginalize over the unknown true time delays and short-scale extrinsic variability. We proceed by describing the generative model in detail.

\subsection{Model for the intrinsic and slow extrinsic variability}

We directly use the free-knot spline fits of our first curve-shifting technique as a model for the intrinsic and slow extrinsic variability. While quasars frequently show variability on scales of hours \citep[see e.g.][]{Stalin:2004iw}, we do not add any additional small-scale variability to the intrinsic spline obtained from the data, since this could exaggeratedly increase the time delay constraining power of our synthetic curves.

We observe that the shapes, amplitudes, and the slopes of the intrinsic and extrinsic splines are virtually insensitive to any plausible time shifts of the light curves around the estimated time delays. Therefore, we simply use the set of splines from our free-knot spline technique as a fixed, deterministic part of our model. We have verified that our results do not significantly change if we perform a time-shift-constrained spline fit on the observed curves for each particular delay that we want to simulate.

Finally, note that the degeneracies between the extrinsic splines used in the model have no influence on the match between the synthetic and observed light curves (see Fig. \ref{fig_comparawsims}). In the scope of our simple generative model we do not need to know whether, for instance, a light curve is amplified by microlensing in image A, or demagnified in B. Only relative extrinsic variability is relevant.

\subsection{Model for the randomized fast extrinsic variability}
\label{noise}

To add fast extrinsic variability to our synthetic light curves, we randomly generate ``power-law noise'', i.e., a time series whose Fourier spectrum follows a power law. The characteristics of this noise are adjusted individually for each quasar image. This procedure can be summarized as follows:

\begin{enumerate}
\item For each quasar light curve, draw power-law noise following \citet{Timmer:1995vt}, using a fine regular sampling of, e.g., 0.2 days.
\item Linearly interpolate these finely sampled signals at the observing epochs of the light curves to obtain a noise contribution for each data point.
\item Locally rescale the noise, so that its amplitude follows the scatter in the observed curves.
\item Run the free-knot spline technique on the synthetic curves, and analyze the residuals.
\item Iteratively repeat the above steps, adjusting the parameters of the power-law noise until the residuals obtained at step 4 are statistically compatible with the residuals obtained from the real observations.
\end{enumerate}

We use this model to generate both correlated extrinsic variability and independent shot noise at once. Therefore, the regular sampling used in step 1 should be chosen significantly finer than the minimal distance between observing epochs.

We now revisit the steps leading to a well-adjusted, fast extrinsic variability in more detail, starting with the power-law noise. The idea of the Timmer \& Koenig algorithm is to generate random amplitude and phase coefficients in the discrete Fourier space, and then build the real signal by inverse Fourier transform. We limit the generation of random amplitudes to a finite window of the frequency domain, thus avoiding any effect on the large-scale variability of our extrinsic model curves. The power-law noise is controlled by the following parameters:
\begin{itemize}
\item $\beta$: the exponent of the spectrum power law; $\beta = -2$ corres\-ponds to a random walk, $\beta = 0$ is white noise);
\item $A$: a scaling factor for the generated noise;
\item $f_{\mathrm{Min}}$: low cut-off of the frequency window.
We set $f_{\mathrm{Min}}$ to $1/500$ day$^{-1}$, as any lower frequencies are by construction well represented by the extrinsic free-knot spline fit;
\item $f_{\mathrm{Max}}$: high cut-off of the frequency window. We take this as the maximum (Nyquist) frequency of the 0.2 day sampling.
\end{itemize}

\begin{figure*}
\resizebox{\hsize}{!}{\includegraphics{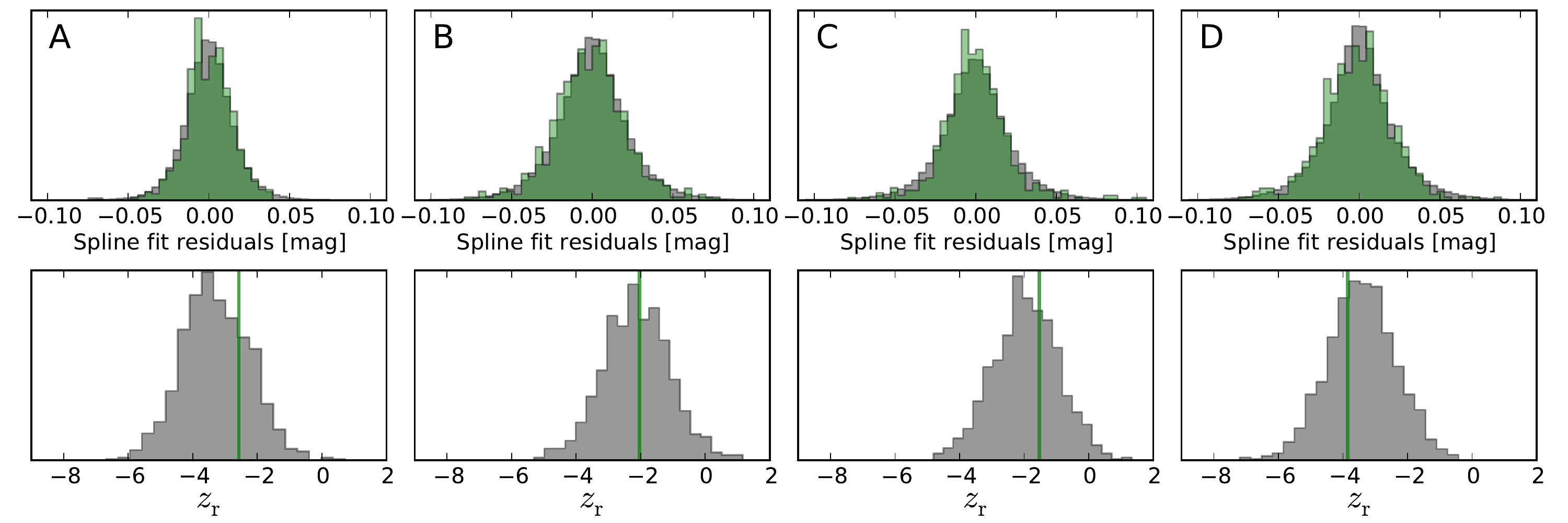}}
\caption{Top: histogram of residuals obtained by running the free-knot spline fit technique on the observed light curves of HE~0435-1223 (green), and the corresponding synthetic curves (gray). Bottom: distribution of the $z_{\mathrm{r}}$ parameter computed from these residuals. Only one set of observed light curves is available, while the distributions related to the synthetic curves are averaged over 1000 realizations.
In this example, the parameters of the generative model have already been adequately adjusted; the synthetic curves shown in Fig. \ref{fig_comparawsims} and \ref{fig_resi} are drawn from this adjusted model.
}
\label{fig_resihist}
\end{figure*}

The free parameters $\beta$ and $A$ have a direct influence on the accuracy and precision that time delay estimators will achieve on the synthetic curves; these are the parameters that will be adjusted for each light curve. We observe that reasonable changes in the parameters $f_{\mathrm{Min}}$ and $f_{\mathrm{Max}}$ (i.e., the sampling) have a negligible effect compared to the influence of $\beta$ and $A$.

Interpolating this power-law noise at the observing epochs leads to a contribution $\epsilon_i$ for each data point. Next, we locally adapt the amplitude of these $\epsilon_i$ to the observed scatter in the light curves, as explained below. A local rescaling is empirically well motivated, because observed light curves often contain subregions that clearly display different smoothness. We compute the rescaling using the residuals $r_{i, \mathrm{obs}}$ of each curve, obtained by running the free-knot spline technique on the observed data (Fig. \ref{fig_resi}). We normalize the absolute values of these noisy residuals to an average of one, and smooth the resulting signal using a median filter with a window of seven observing epochs:
\begin{equation}
s_i = \textrm{median} \left( \frac{|r_{j, \mathrm{obs}}|}{\langle| r_{\mathrm{obs}} |\rangle}, j \in \{i-3, \ldots, i+3\} \right),
\end{equation}
where $\langle| r_{\mathrm{obs}} |\rangle$ denotes the average absolute residual taken over all points of the curve.
The rescaled $\epsilon_i$ that we add to our synthetic light curves are given by
\begin{equation}
\epsilon_{i, \mathrm{rescale}} = \epsilon_i \cdot s_i.
\end{equation}
This local rescaling does not affect the average amplitude of the synthetic noise, which is still controlled by $A$.

We proceed by running the spline-fitting technique from scratch on the set of synthetic curves, i.e., disregarding any information from the generative model. This yields residuals, shown as black points in Fig. \ref{fig_resi}, that can be equitably compared with the residuals of the observations.

To fine-tune $A$ and $\beta$, we quantitatively analyze these residuals. For this, we make use of two simple statistics: their standard deviation $\sigma$ and their number of ``runs'' $r$. A \emph{run} is a sequence of adjacent residuals that are all either positive or negative. The total number of positive and negative runs is a statistic that can be used to test the hypothesis that successive residuals are independent \citep[][chap. 5]{Wall:2003ud}. For large samples of $N$ truly independent observations with $N_{+}$ positive and $N_{-}$ negative residuals, the number of runs $r$ is normally distributed with an expectation value and a variance respectively given by 
\begin{equation}
\mu_{r} = \frac{2 N_{+} N_{-}}{N} + 1 \quad \textrm{and} \quad \sigma_{r}^2 = \frac{ (\mu_{r} - 1) (\mu_{r} - 2) }{N - 1} .
\end{equation}
In practice we ``normalize'' a measured number of runs $r$ to this hypothesis of independent residuals:
\begin{equation}
z_{r} = \frac{r - \mu_{r}}{\sigma_{r}}.
\end{equation}

Applying the free-knot spline technique on COSMOGRAIL light curves, we typically observe a number of runs in the residuals at least about $2\sigma_{r}$ lower than $\mu_{r}$, i.e., $z_{r} \approx -2$, meaning that the residuals are indeed correlated.
Figure \ref{fig_resihist} shows a comparison in terms of residual distribution and $z_{r}$ between the residuals obtained from synthetic curves (gray distributions, averaged over 1000 realizations) and from the observations of HE~0435-1223.
One can now adjust the parameters $A$ and $\beta$ of the power-law noise so that the average standard deviation and number of runs of the residuals obtained from the synthetic curves match the values measured on the observations. The procedure is straightforward because the residuals' number of runs directly relates to $\beta$, and their standard deviation to $A$, without much crosstalk. A good match such as shown in Fig. \ref{fig_resihist} is thus obtained after just a few trial-and-error iterations, simultaneously modifying $A$ and $\beta$ of each quasar image.

Finally, we attribute the photometric error bars of the observed data to all our synthetic light curves. We stress that we do not make use of these photometric error bars anywhere else in our generative model. These photometric error bars only describe the experimental measurement uncertainty, which is not necessarily the only source of short-scale ``mismatch'' between multiple light curves. Furthermore, for all methods presented in this paper, the photometric error bars essentially only act as \emph{relative weights} between the data points. As a consequence, our time delay measurements -- including their uncertainty estimates -- are not directly influenced by a potential systematic under- or over-estimation of the photometric errors. We see this as a very desirable feature.

\subsection{Quantifying variance and bias of our time delay estimators}
\label{hists}

\begin{figure*}[htbp]
\resizebox{\hsize}{!}{\includegraphics{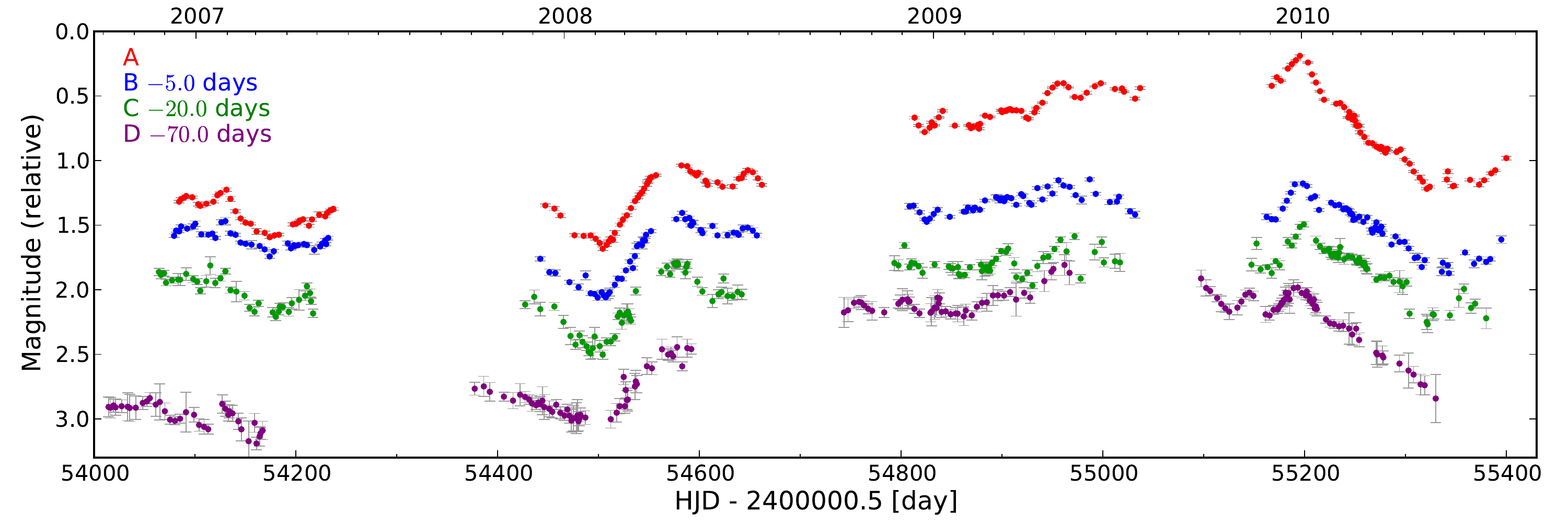}}
\caption{The ``trial curves'', i.e., a set of artificial 4-season long light curves of a quad lens, used in place of real observations for the self-consistency check performed in Section \ref{test}. These curves are drawn from a generative model that closely mimics COSMOGRAIL observations. Note the prominent microlensing on large time scales, but also the presence of obviously extrinsic variability on scales of weeks (e.g., middle of third season of curve C). The true time delays are $\Delta t_{\mathrm{AB}} = -5.0$, $\Delta t_{\mathrm{AC}} = -20.0$, and $\Delta t_{\mathrm{AD}} = -70.0$ days. In this figure the curves are shifted according to these delays.}
\label{fig_artificial}
\end{figure*}

With the well-adjusted generative model in hand, we proceed by drawing typically 1000 synthetic curve sets, choosing model time shifts in a range of several days around the point estimates obtained from the observations. We run the curve-shifting technique on these synthetic curves, using the same parameters as for the estimation performed on the real observations. We always start the techniques from random initial time shifts, to take the potential intrinsic variance of the curve-shifting technique described in Section \ref{techniques} into account. We then compute the resulting errors between the estimated and the true time delays. 

To analyze these errors, we bin the synthetic curve sets according to their true time \emph{delays} individually for each quasar image pair. This allows us to check if the uncertainties that we derive do not strongly depend on the true time delays. Figure \ref{fig_artificial_meanvstrue} illustrates the procedure in the case of a quad lens. In each bin, we estimate the \emph{variance} and the \emph{bias} of our time delay estimators, by computing the sample variance $(\sigma^2_{\mathrm{ran}})$ and the average $(\sigma_{\mathrm{sys}})$ of the delay measurement errors, respectively.

By construction, the main origin of the estimated bias has to be related to those properties of light curves that are kept constant in the Monte Carlo simulations, such as the interplay between intrinsic and slow extrinsic variability, season gaps, and sampling. This bias quantifies the accuracy of a time delay estimation. Such a characterization of accuracy has often been neglected in past time delay measurements; it cannot be obtained through resampling techniques that do not involve a generative model, such as jackknifing or bootstrapping.

On the other hand, the variance of our time delay estimators mainly results from their sensitivity to the fast extrinsic variability and noise, which we randomize in our synthetic curves. This variance describes the precision of the time delay measurements. It is often possible to increase the precision of a curve-shifting technique, by somehow smoothing either the input light curves or the cost function to be optimized. Using the approach described in this paper, we can verify that such an increase in precision is not obliterated by an even higher decrease in accuracy.

Lastly, we obtain a comprehensive one-sigma error bar $\sigma_{\mathrm{tot}}$ for the time delay measurement between each quasar image pair by combining the maximal bias and the maximal variance observed in the bins of true time delays:
\begin{equation}
\label{sigma}
\sigma_{\mathrm{tot}} = \sqrt{\,\max_{\Delta t_{\mathrm{True}}} \, \sigma^2_{\mathrm{ran}} + \max_{\Delta t_{\mathrm{True}}} \,\sigma^2_{\mathrm{sys}}}.
\end{equation}

In situations where bias and variance do not significantly depend on the true time delays, $\sigma_{\mathrm{tot}}^2$ simply corresponds to the mean squared error (MSE) of our time delay estimators. The analysis shown in Fig. \ref{fig_artificial_meanvstrue} roughly corresponds to such a situation.

As becomes evident in the next section, it appears very tempting to empirically \emph{correct} each time delay estimation for its bias, instead of simply combining the bias and the variance into the total error budget. To avoid any circular argumentation, we do not perform such a correction in this paper. The circularity would arise since we have to assume time delays for the observed data in order to \emph{build} the generative model of the synthetic curves. If these initial guesses for time delays are significantly biased, the generative model will contain erroneous intrinsic and extrinsic variability patterns, yet still mimic the observed curves. Running the curve-shifting techniques on the synthetic curves would therefore yield measured delays close to their true delays, and thus not fully reveal the initial errors. 

We prefer to simply use the amplitude of the bias, gauged through the above procedure, as a quality criterion of each method.

\section{Application to trial curves and discussion}
\label{test}

In this section, we apply our three curve-shifting techniques, as well as the error bar computation procedure of Section \ref{errorbars}, to a set of artificial curves with known time delays. This allows us to check the consistency of the error bar computation, and to illustrate some general observations about the latter.

\subsection{The trial curves}

We use the scheme described in Section \ref{errorbars} to generate a single set of four-season-long light curves that mimic the data of one of the quadruply imaged quasars monitored by the COSMOGRAIL collaboration. In the following, we refer to these artificial curves as the ``trial curves'', shown in Fig. \ref{fig_artificial}. The trial curves include realistic intrinsic variability and sampling, as well as obvious long- and short-scale microlensing. We choose time delays of $\Delta t_{\mathrm{AB}} = -5.0$, $\Delta t_{\mathrm{AC}} = -20.0$, and $\Delta t_{\mathrm{AD}} = -70.0$ days.

\begin{figure*}[ht!]
\begin{center}
\vskip 35pt
\resizebox{0.8 \hsize}{!}{\includegraphics{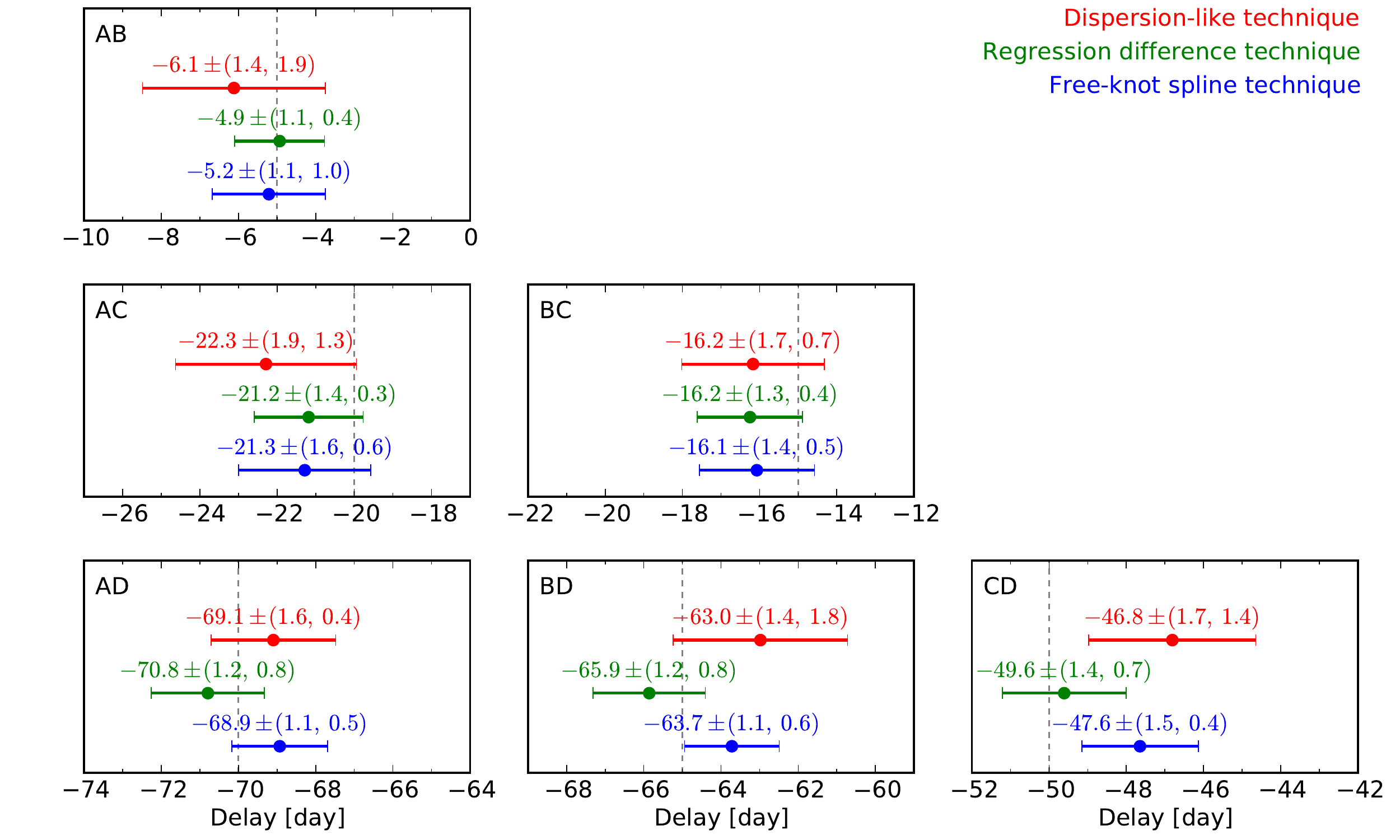}}
\caption{Time delay estimations and associated uncertainties obtained by each of the three curve-shifting techniques from the trial curves in Fig. \ref{fig_artificial}. For each delay measurement, the random error bar, $\sigma_{\mathrm{ran}}$ and the bias, $\sigma_{\mathrm{sys}}$, are given in parentheses. The drawn error bars depict the total error, $\sigma_{\mathrm{tot}}$, as obtained from equation \ref{sigma}. The dashed vertical lines show the true delays of the trial curves analyzed in this section.}
\label{fig_artificial_delays}
\vskip 25pt
\resizebox{0.8 \hsize}{!}{\includegraphics{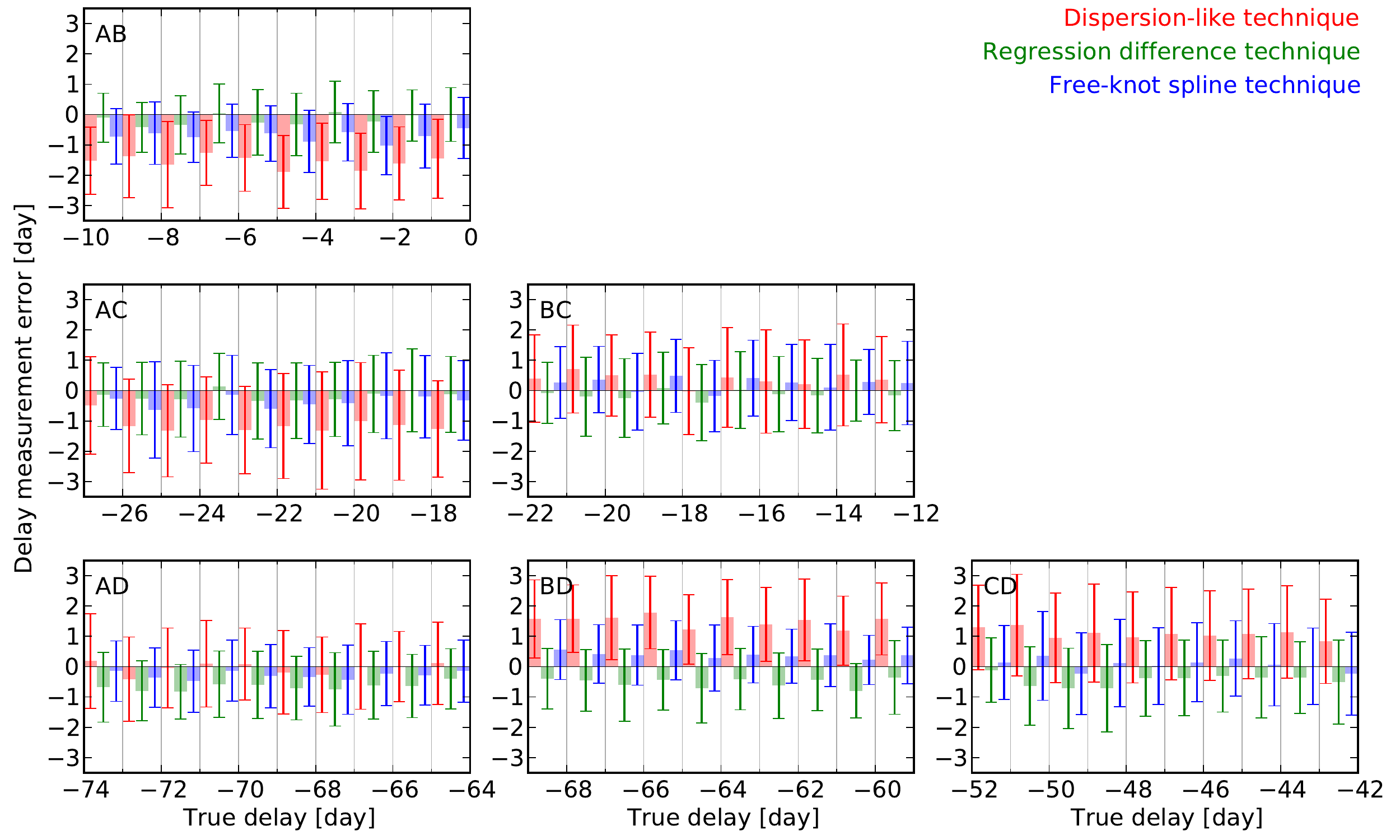}}
\caption{Error analysis for the trial curves shown in Fig. \ref{fig_artificial}. For each curve-shifting technique, the estimates of the time delay uncertainties are obtained by analyzing delay measurement errors on 1000 synthetic curves with randomly chosen true time delays. In each panel, the delay measurement errors (vertical axis) are represented against the true delays of these synthetic curves (horizontal axis). Instead of showing a scatter plot, the averaged measurement error (i.e., the bias $\sigma_{\mathrm{sys}}$) in bins of true delay is shown by the shaded rods, while the error bars represent the standard deviation $\sigma_{\mathrm{ran}}$. The bin intervals are indicated by the light vertical lines. For increased clarity, the results from the three techniques are drawn side by side within each bin.}
\label{fig_artificial_meanvstrue}
\end{center}
\end{figure*}

\subsection{Analysis}

From here on we process the trial curves as if they were real observations. We use our time delay estimators on them, disregarding any prior information about the true delays and intrinsic or extrinsic variations. For the spline method, we choose an average knot step of 20 days for the intrinsic spline, and 150 days for the four extrinsic splines. The dispersion-like method is allowed to correct for extrinsic variability using independent linear trends on each of the 16 seasons. All other parameters of the curve-shifting techniques are fixed to the default values presented earlier.

As described in Section~\ref{techniques}, we systematically run the curve-shifting methods several hundreds of times on the same light curves, starting from randomized initial time shifts, and we use the mean of the resulting time delay distributions as our best estimation. This leads to the centroids of the time delay estimates, presented in Fig. \ref{fig_artificial_delays}.

Finally, we compute error bars for these time delay measurements following the procedure of Section~\ref{errorbars}. In doing so, the curves used for the Monte Carlo runs are drawn from a {\it new generative model built from scratch} from the trial curves. Note that despite this precaution, the same kind of recipes were used to build the trial curves and the synthetic curves used in the Monte Carlo procedure. The analysis performed in this section should therefore be seen as a self-consistency check of our time delay uncertainty estimation procedure. 

We present this error analysis in Fig. \ref{fig_artificial_meanvstrue}. For each bin of true delay used in the Monte Carlo simulations, we show the mean $\sigma_{\mathrm{sys}}$ and standard deviation $\sigma_{\mathrm{ran}}$ of the measurement errors as a shaded rod and as an associated error bar, respectively. To give an example, we can observe on this figure that the dispersion-like technique systematically underestimates the delay $\Delta t_{\mathrm{AB}}$ by about 1.5 days, regardless of the true delay used in the synthetic curves. For this specific curve-shifting technique and pair of curves, the bias is larger than the standard deviation of the measurement errors, hence underlining the importance of evaluating the bias.

The Monte Carlo procedure described in Section \ref{errorbars} allows us to perform the analysis summarized in Fig. \ref{fig_artificial_meanvstrue} for any set of observed light curves. From such an analysis, we directly obtain the total uncertainty $\sigma_{\mathrm{tot}}$ following Eq.~\ref{sigma}, for each technique and pair of quasar images. These $\sigma_{\mathrm{tot}}$ are represented as error bars in Fig. \ref{fig_artificial_delays}.

\subsection{Discussion}

Several important observations can be made from Figs.~\ref{fig_artificial_delays} \& \ref{fig_artificial_meanvstrue}. We recall that the analysis leading to these figures did not use any knowledge of the true time delays between the trial curves at any time, except for plotting the dashed vertical lines in Fig. \ref{fig_artificial_delays}.

\begin{enumerate}

\item We observe in Fig. \ref{fig_artificial_delays} that, on average, the total 1$\sigma$ error bars computed by our Monte Carlo approach compare well with the actual errors made by our point estimators on the trial curves. In particular, this holds for the dispersion-like technique, which suffers in this example from the largest biases. The self-consistency check of our uncertainty estimation procedure is successful.

\item Figure \ref{fig_artificial_meanvstrue} clearly shows that, in the case of these trial curves, our estimates of both the bias and the variance of each curve-shifting method does not depend much on the true time delays of the Monte Carlo simulations. This optimal situation is often not observed for shorter or lower quality curves, motivating our conservative decision to combine the maximum bias and variance to get the total error bar following equation \ref{sigma}.

\item
We can make an additional observation about the \emph{direction and magnitude} of the bias. Consider for example the time delay $\Delta t_{\mathrm{BD}}$. From Fig.~\ref{fig_artificial_delays} we see that the lowest estimate of the delay is obtained by the regression difference technique, followed by the spline technique, and then by the dispersion-like technique. 
Independently, Fig. \ref{fig_artificial_meanvstrue} shows that on our Monte Carlo simulations, the regression difference technique tends to underestimate this delay, the spline technique tends to slightly overestimates it, while the dispersion-like technique overestimates it on average by $\sim1.5$ days. This means that the ``order'' of the biases as estimated from the Monte Carlo simulations is consistent with the sequence of time delays measured on the trial curves. A similar statement holds for nearly all pairs of quasar images, which is remarkable. We conclude that our uncertainty estimation procedure of Section \ref{errorbars} is at least in part successful in separating the bias due to flaws of the methods from the random error that genuinely comes from the data.
This aspect can be analyzed for real data, as it does not involve knowledge of the true time delays between the trial curves.

\end{enumerate}

\subsection{Investigating error correlations between pairs of quasar images}
\label{correlations}

The error bars computed for a given delay measurement with our method marginalize over the other delays of the same lens. However, the delay measurements are not independent in their very nature. For example, $\Delta t_{\mathrm{AC}} = \Delta t_{\mathrm{AB}} + \Delta t_{\mathrm{BC}}$, hence any estimation of $\Delta t_{\mathrm{AC}}$ correlates positively with both $\Delta t_{\mathrm{AB}}$ and $\Delta t_{\mathrm{BC}}$. The formalism presented in this paper does not fully exploit this joint information.

\begin{figure}[tbp]
\resizebox{\hsize}{!}{\includegraphics{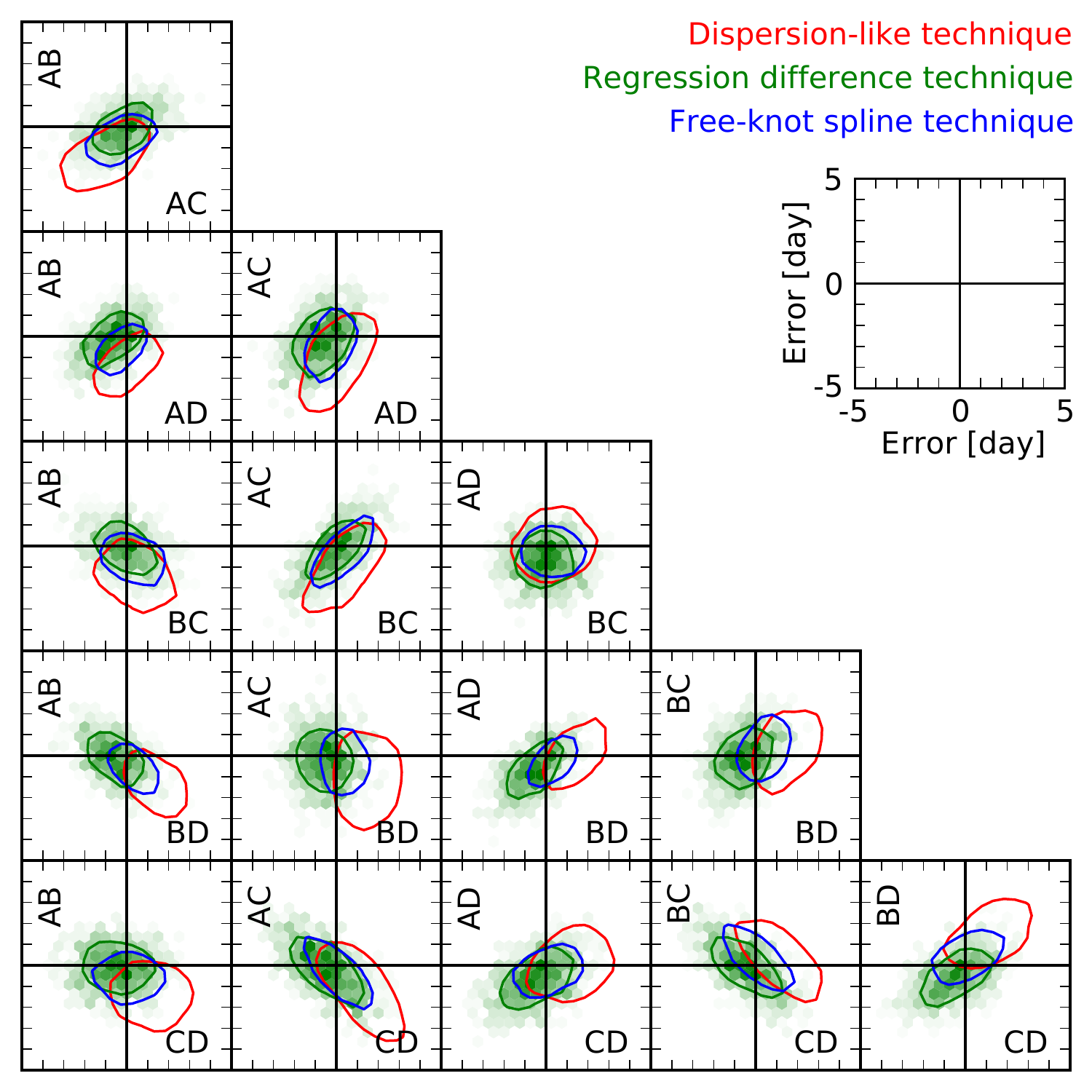}}
\caption{Correlations of delay measurement errors for the quad lens analyzed in Section \ref{test}. The distribution of delay measurement errors on 1000 synthetic curves is shown for each quasar image pair against each other pair, marginalizing over the true delays of the synthetic curves. The central crosshair of each panel indicates zero error, and the ticks are days; i.e., each axis shows errors from $-5$ to $+5$ days. Displacements of the distributions with respect to the crosshairs indicate bias, while the width of the distributions indicates variance of the time delay estimators. For clarity, only single contours at half of the maximum density are shown for two of the techniques. The measurement errors shown in this figure are exactly the same as used in Fig. \ref{fig_artificial_meanvstrue}.
}
\label{fig_artificial_cov}
\end{figure}

We can nevertheless use the measurements obtained from the Monte Carlo simulations to explore correlations between delay estimation errors of different pairs of quasar images. In Fig. \ref{fig_artificial_cov}, we show this correlation for pairs of curves, marginalizing over the true delays used in the Monte Carlo simulations.

Displacements of these distributions with respect to the zero-error crosshairs indicate bias, while the width of the distributions indicates the variance of the time delay estimators. As expected, we observe positive and negative correlations, which are ellipsoids with oblique orientations, for pairs of delays that involve a common quasar image. Very importantly, at least for the specific set of light curves used in this work, we do \emph{not} observe correlations between the three ``disjoint'' pairs of delays AB/CD, AC/BD, and AD/BC, located along the short diagonal of Fig.~\ref{fig_artificial_cov}. 

This analysis is to be done for the light curves of every specific quadruply imaged quasar, since the sensitivity of the curve-shifting algorithms certainly depends on the details of the features in real quasar light curves. If the results are qualitatively similar to those shown in Fig. \ref{fig_artificial_cov}, we deem it appropriate to present our time delay measurements in the form of independent estimates. In practice a joint probability distribution is only really interesting for quadruply imaged quasars in which several delays are well measurable.

\subsection{Getting a single answer}
\label{combination}

It remains to be decided how to ``combine'' the delay measurements obtained from our different techniques. We cannot combine the measurements as if they were independent, for instance by multiplying associated probability density functions. Indeed, no matter how different the \emph{methods} are, they all use the one and only realization of the observations. In particular, it can well happen that the time delay uncertainties, as derived by our approach for each method, are significantly larger than the apparent spread of the delay point estimates obtained from the different methods (see e.g. panel BC of Fig. \ref{fig_artificial_delays}). The spread of different estimates is clearly not an indication of the degree of knowledge of a time delay.

As a result, when analyzing real observations, we propose to simply select, individually for each lens, the method that tends to display the smallest total uncertainty. One can directly use its point estimates and the associated $1\sigma$ error bars as the favored answer.

Average delays between competitive methods could also be used, but in any case the size of the uncertainties are not to be divided by the square root of the number of methods contributing to these averages.

\section{Conclusions}
\label{conclusions}

In this paper, we describe three independent ``curve-shifting techniques'' to measure time delays between resolved light curves of gravitationally lensed quasars. All these methods address the presence of variable microlensing in the light curves and can be applied to lens systems with any number of images.
\begin{enumerate}
\item {\bf The free-knot spline technique} simultaneously fits one common \emph{intrinsic} spline and independent smoother \emph{extrinsic} splines to the light curves. The curves are shifted in time so as to optimize this fit.
\item {\bf The regression difference technique} shifts regressions of the curves to minimize the \emph{variability} of the differences between them. It is nearly parameter free and does not require an explicit model for the microlensing variability.
\item {\bf The dispersion-like technique} shifts the curves so as to minimize a measure of the \emph{dispersion} between the overlapping data points. This method has no explicit model for the common intrinsic variability of the quasar, but it involves polynomial models for the extrinsic variability. It has previously been applied in \citet{Courbin:2011bl}.
\end{enumerate}
A common point of the methods is that they yield point estimates (i.e., single values) for the time delays in a self-consistent way by sharing the formalism of time \emph{shifts} described in Section \ref{techniques}.

In addition, we present a Monte Carlo approach to estimate the uncertainty of each time delay measurement, including both random and systematic errors. This procedure is based on synthetic curves that try to mimic as much information about the intrinsic and extrinsic variability as the observations unmistakenly reveal. Provided that we accept the generative model of the synthetic curves, the curve-shifting techniques themselves are reduced to ``recipes''. Given a set of light curves, we can select methods based solely upon their empirical performance. This effectively shifts the requirement of formal justification from the curve-shifting techniques to the synthetic curves on which these techniques are evaluated. As a consequence, the techniques can even be fine tuned for each data set.

Finally, we verified the self-consistency of our time delay uncertainty estimation using a trial set of artificial light curves. The availability of three different curve-shifting techniques allows consistency checks of our \emph{bias} determination to be performed when analyzing real observations, i.e., data without known time delays. In other words, we acknowledge that any curve-shifting technique may display residual biases due, for example, to particular patterns of slow microlensing variability, but we provide a means to evaluate this bias.

The methods described in this paper will be used as a standard benchmark to obtain time delay estimations from the light curves of the COSMOGRAIL monitoring program. We implemented the curve-shifting techniques, the error bar estimation, and the generation of all the figures of this paper in the form of a modular \verb+python+ toolbox. This package, called PyCS, as well as a tutorial including the trial curves of Section \ref{test} are available from the COSMOGRAIL website\footnote{\url{http://www.cosmograil.org}}.

\begin{acknowledgements}
This work is supported by the Swiss National Science Foundation (SNSF). We acknowledge support from the International Space Science Institute where this research was initiated, and thank Eva Eulaers, Pierre Magain, and the anonymous referee for helpful discussions and comments. It is also a pleasure to acknowledge heavy use of the \verb+scipy+ \citep{numpyscipy} and \verb+matplotlib+ \citep{matplotlib} projects, which are both community efforts.
\end{acknowledgements}

\bibliographystyle{aa} 
\bibliography{papers} 

\begin{thebibliography}{45}
\expandafter\ifx\csname natexlab\endcsname\relax\def\natexlab#1{#1}\fi

\bibitem[{Barkana(1997)}]{Barkana:1997cv}
Barkana, R. 1997, ApJ, 489, 21

\bibitem[{Burud {et~al.}(2001)Burud, Magain, Sohy, \& Hjorth}]{Burud:2001ka}
Burud, I., Magain, P., Sohy, S., \& Hjorth, J. 2001, A{\&}A, 380, 805

\bibitem[{Courbin {et~al.}(2011)Courbin, Chantry, Revaz, Sluse, Faure, Tewes,
  Eulaers, Koleva, Asfandiyarov, Dye, Magain, van Winckel, Coles, Saha,
  Ibrahimov, \& Meylan}]{Courbin:2011bl}
Courbin, F., Chantry, V., Revaz, Y., {et~al.} 2011, A{\&}A, 536, A53

\bibitem[{Cuevas-Tello {et~al.}(2006)Cuevas-Tello, Tino, \&
  Raychaudhury}]{CuevasTello:2006fn}
Cuevas-Tello, J.~C., Tino, P., \& Raychaudhury, S. 2006, Lecture Notes in
  Computer Science, 4212, 614

\bibitem[{Cuevas-Tello {et~al.}(2010)Cuevas-Tello, Tino, Raychaudhury, Yao, \&
  Harva}]{CuevasTello:2010td}
Cuevas-Tello, J.~C., Tino, P., Raychaudhury, S., Yao, X., \& Harva, M. 2010,
  Pattern Recognition, 43, 1165

\bibitem[{de~Boor(1978)}]{deBoor:1978wq}
de~Boor, C. 1978, {A Practical Guide to Splines} (Springer)

\bibitem[{Dierckx(1995)}]{Dierckx:1995tp}
Dierckx, P. 1995, {Curve and Surface Fitting With Splines} (Clarendon Press)

\bibitem[{Eigenbrod {et~al.}(2008{\natexlab{a}})Eigenbrod, Courbin, Meylan,
  Agol, Anguita, Schmidt, \& Wambsganss}]{Eigenbrod:2008dw}
Eigenbrod, A., Courbin, F., Meylan, G., {et~al.} 2008{\natexlab{a}}, A{\&}A,
  490, 933

\bibitem[{Eigenbrod {et~al.}(2008{\natexlab{b}})Eigenbrod, Courbin, Sluse,
  Meylan, \& Agol}]{Eigenbrod:2008iq}
Eigenbrod, A., Courbin, F., Sluse, D., Meylan, G., \& Agol, E.
  2008{\natexlab{b}}, A{\&}A, 480, 647

\bibitem[{Eigenbrod {et~al.}(2005)Eigenbrod, Courbin, Vuissoz, Meylan, Saha, \&
  Dye}]{Eigenbrod:2005jx}
Eigenbrod, A., Courbin, F., Vuissoz, C., {et~al.} 2005, A{\&}A, 436, 25

\bibitem[{Eulaers \& Magain(2011)}]{Eulaers:2011hz}
Eulaers, E. \& Magain, P. 2011, A{\&}A, 536, A44

\bibitem[{Gil-Merino {et~al.}(2002)Gil-Merino, Wisotzki, \&
  Wambsganss}]{GilMerino:2002fi}
Gil-Merino, R., Wisotzki, L., \& Wambsganss, J. 2002, A{\&}A, 381, 428

\bibitem[{Golub \& Pereyra(1973)}]{Golub:1973wb}
Golub, G. \& Pereyra, V. 1973, SIAM Journal on numerical analysis, 413

\bibitem[{Harva \& Raychaudhury(2008)}]{Harva:2008wy}
Harva, M. \& Raychaudhury, S. 2008, Neurocomputing, 72, 32

\bibitem[{Hirv {et~al.}(2011)Hirv, Olspert, \& Pelt}]{Hirv:2011wt}
Hirv, A., Olspert, N., \& Pelt, J. 2011, Baltic Astronomy, 20, 125

\bibitem[{Hunter(2007)}]{matplotlib}
Hunter, J.~D. 2007, Computing in Science {\&} Engineering, 9, 90

\bibitem[{Kochanek(2004)}]{Kochanek:2004ir}
Kochanek, C.~S. 2004, ApJ, 605, 58

\bibitem[{Kochanek {et~al.}(2007)Kochanek, Dai, Morgan, Morgan, \&
  Poindexter}]{Kochanek:2007tz}
Kochanek, C.~S., Dai, X., Morgan, C., Morgan, N., \& Poindexter, S. C.~G. 2007,
  in Statistical Challenges in Modern Astronomy IV, ASP Conference Series, 371

\bibitem[{Kochanek {et~al.}(2006)Kochanek, Morgan, Falco, Mcleod, Winn,
  Dembicky, \& Ketzeback}]{Kochanek:2006fp}
Kochanek, C.~S., Morgan, N.~D., Falco, E.~E., {et~al.} 2006, ApJ, 640, 47

\bibitem[{Kundic {et~al.}(1997)Kundic, Turner, Colley, Gott~III, Rhoads, Wang,
  Bergeron, Gloria, Long, Malhotra, \& Wambsganss}]{Kundic:1997br}
Kundic, T., Turner, E.~L., Colley, W.~N., {et~al.} 1997, ApJ, 482, 75

\bibitem[{Lehar {et~al.}(1992)Lehar, Hewitt, Burke, \& Roberts}]{Lehar:1992bg}
Lehar, J., Hewitt, J.~N., Burke, B.~F., \& Roberts, D.~H. 1992, ApJ, 384, 453

\bibitem[{Linder(2011)}]{Linder:2011cs}
Linder, E.~V. 2011, Physical Review D, 84, 123529

\bibitem[{Molinari {et~al.}(2004)Molinari, Durand, \&
  Sabatier}]{Molinari:2004tv}
Molinari, N., Durand, J., \& Sabatier, R. 2004, Computational statistics {\&}
  data analysis, 45, 159

\bibitem[{Morgan {et~al.}(2008)Morgan, Eyler, Kochanek, Morgan, Falco, Vuissoz,
  Courbin, \& Meylan}]{Morgan:2008fc}
Morgan, C.~W., Eyler, M.~E., Kochanek, C.~S., {et~al.} 2008, ApJ, 676, 80

\bibitem[{Morgan {et~al.}(2012)Morgan, Hainline, Chen, Tewes, Kochanek, Dai,
  Kozlowski, Blackburne, Mosquera, Chartas, Courbin, \& Meylan}]{Morgan:2012cl}
Morgan, C.~W., Hainline, L.~J., Chen, B., {et~al.} 2012, ApJ, 756, 52

\bibitem[{Oliphant(2007)}]{numpyscipy}
Oliphant, T. 2007, Computing in Science {\&} Engineering, 9, 10

\bibitem[{Patil {et~al.}(2010)Patil, Huard, \& Fonnesbeck}]{Patil:2010tx}
Patil, A., Huard, D., \& Fonnesbeck, C. 2010, Journal of statistical software,
  35, 1

\bibitem[{Pelt {et~al.}(1998)Pelt, Hjorth, Refsdal, Schild, \&
  Stabell}]{Pelt:1998vc}
Pelt, J., Hjorth, J., Refsdal, S., Schild, R., \& Stabell, R. 1998, A{\&}A,
  337, 681

\bibitem[{Pelt {et~al.}(1996)Pelt, Kayser, Refsdal, \& Schramm}]{Pelt:1996vy}
Pelt, J., Kayser, R., Refsdal, S., \& Schramm, T. 1996, A{\&}A, 305, 97

\bibitem[{Pelt {et~al.}(2002)Pelt, Refsdal, \& Stabell}]{Pelt:2002ke}
Pelt, J., Refsdal, S., \& Stabell, R. 2002, A{\&}A, 389, L57

\bibitem[{Press {et~al.}(1992)Press, Rybicki, \& Hewitt}]{Press:1992jj}
Press, W.~H., Rybicki, G.~B., \& Hewitt, J.~N. 1992, ApJ, 385, 404

\bibitem[{Press {et~al.}(2007)Press, Teukolsky, Vetterling, \&
  Flannery}]{Press:2007tn}
Press, W.~H., Teukolsky, S.~A., Vetterling, W.~T., \& Flannery, B.~P. 2007,
  {Numerical Recipes Third Edition} (Cambridge University Press)

\bibitem[{Refsdal(1964)}]{Refsdal:1964vh}
Refsdal, S. 1964, MNRAS, 128, 307

\bibitem[{Refsdal {et~al.}(2000)Refsdal, Stabell, Pelt, \&
  Schild}]{Refsdal:2000tm}
Refsdal, S., Stabell, R., Pelt, J., \& Schild, R. 2000, A{\&}A, 360, 10

\bibitem[{Schechter {et~al.}(2003)Schechter, Udalski, Szyma{\'n}ski, Kubiak,
  Pietrzynski, Soszynski, Wo{\'z}niak, {\.Z}ebru{\'n}, Szewczyk, Wyrzykowski,
  \& {The OGLE Collaboration}}]{Schechter:2003ep}
Schechter, P.~L., Udalski, A., Szyma{\'n}ski, M., {et~al.} 2003, ApJ, 584, 657

\bibitem[{Stalin {et~al.}(2005)Stalin, Gupta, Gopal-Krishna, Wiita, \&
  Sagar}]{Stalin:2004iw}
Stalin, C.~S., Gupta, A.~C., Gopal-Krishna, Wiita, P.~J., \& Sagar, R. 2005,
  MNRAS, 356, 607

\bibitem[{Suyu {et~al.}(2010)Suyu, Marshall, Auger, Hilbert, Blandford,
  Koopmans, Fassnacht, \& Treu}]{Suyu:2010fq}
Suyu, S.~H., Marshall, P.~J., Auger, M.~W., {et~al.} 2010, ApJ, 711, 201

\bibitem[{Suyu {et~al.}(2009)Suyu, Marshall, Blandford, Fassnacht, Koopmans,
  McKean, \& Treu}]{Suyu:2009ig}
Suyu, S.~H., Marshall, P.~J., Blandford, R.~D., {et~al.} 2009, ApJ, 691, 277

\bibitem[{Timmer \& Koenig(1995)}]{Timmer:1995vt}
Timmer, J. \& Koenig, M. 1995, A{\&}A, 300, 707

\bibitem[{Vakulik {et~al.}(2009)Vakulik, Shulga, Schild, Tsvetkova, Dudinov,
  Minakov, Nuritdinov, Artamonov, Kochetov, Smirnov, Sergeyev, Konichek,
  Sinelnikov, Bruevich, Akhunov, \& Burkhonov}]{Vakulik:2009br}
Vakulik, V.~G., Shulga, V.~M., Schild, R.~E., {et~al.} 2009, MNRAS, 400, L90

\bibitem[{Vanderriest {et~al.}(1989)Vanderriest, Schneider, Herpe, Chevreton,
  Moles, \& Wlerick}]{Vanderriest:1989uj}
Vanderriest, C., Schneider, J., Herpe, G., {et~al.} 1989, A{\&}A, 215, 1

\bibitem[{Vuissoz {et~al.}(2008)Vuissoz, Courbin, Sluse, Meylan, Chantry,
  Eulaers, Morgan, Eyler, Kochanek, Coles, Saha, Magain, \&
  Falco}]{Vuissoz:2008bu}
Vuissoz, C., Courbin, F., Sluse, D., {et~al.} 2008, A{\&}A, 488, 481

\bibitem[{Wall \& Jenkins(2003)}]{Wall:2003ud}
Wall, J. \& Jenkins, C. 2003, {Practical Statistics for Astronomers} (Cambridge
  University Press)

\bibitem[{Walsh {et~al.}(1979)Walsh, Carswell, \& Weymann}]{Walsh:1979cz}
Walsh, D., Carswell, R.~F., \& Weymann, R.~J. 1979, Nature, 279, 381

\bibitem[{Wambsganss {et~al.}(2000)Wambsganss, Schmidt, Colley, Kundi{\'c}, \&
  Turner}]{Wambsganss:2000uc}
Wambsganss, J., Schmidt, R.~W., Colley, W., Kundi{\'c}, T., \& Turner, E.~L.
  2000, A{\&}A, 362, L37

\end{thebibliography}
\end{document}